\documentclass[12pt]{article}

\usepackage{bibentry}          
\usepackage{attrib}            
\usepackage{bm}
\usepackage{setspace}
\usepackage{graphicx,scalerel}
\usepackage{amsmath}
\usepackage{array}
\usepackage{url}
\usepackage{amssymb}
\usepackage{caption}
\usepackage{float}
\usepackage{enumitem} 
\usepackage{subcaption}
\usepackage{natbib}
\usepackage{caption}
\usepackage{framed}
\usepackage{multirow} 
\usepackage{setspace}
\usepackage{color}
\usepackage{tikz}
\usepackage{pgfgantt}
\usepackage{mathtools}
\usepackage{nccmath}
\usepackage{longtable}
\usepackage[utf8]{inputenc} 
\usepackage[T1]{fontenc} 
\usepackage{orcidlink,thumbpdf,lmodern}
\usepackage{framed}
\usepackage{fancyvrb}
\PassOptionsToPackage{colorlinks=true, allcolors=black}{hyperref}
\usepackage{hyperref}
\hypersetup{
    hidelinks,
    colorlinks=false,
    pdfborder={0 0 0}
}
\usepackage{parskip}
\AtBeginEnvironment{longtable}{\addtocounter{table}{-1}\refstepcounter{table}}

\makeatletter
\newcommand\notsotiny{\@setfontsize\notsotiny{7.5}{9.5}}
\makeatother

\usepackage{geometry}
\begin{document}

\begin{center}
{\Large{\textbf{\texttt{hassediagrams}: an \texttt{R} package that generates the Hasse diagram of the layout structure and the restricted layout structure}} } \\[1ex] 
		
Damianos Michaelides\footnote{Damianos Michaelides; Current affiliation: Biostatistics Unit, The Cyprus Institute of Neurology and Genetics; damianosm@cing.ac.cy}
,Simon T. Bate \& Marion J. Chatfield \\[1ex]
		
GlaxoSmithKline \\
United Kingdom
\end{center}			
				

\normalsize

With the advent of modern statistical software, complex experimental designs are now routinely employed in many areas of research. Failing to correctly identify the structure of the experimental design can lead to incorrect model selection and misleading inferences. This paper describes the \texttt{hassediagrams} package in \texttt{R} that determines the structure of the design, summarised by the \textit{layout structure}, and generates a Hasse diagram of the layout structure. By considering the randomisation performed, in conjunction with the layout structure, a set of randomisation objects can be defined that form the \textit{restricted layout structure}. This structure can also be visualised using a generalisation of the Hasse diagram. Objects in the restricted layout structure can be used to identify the terms to include in the statistical model. The use of the procedure thus ensures consistency of model selection due to the systematic approach taken to generate the model.

\begin{flushleft}
Keywords: Hasse diagrams, layout structure, restricted layout structure, randomisation, experimental designs in \texttt{R}
\end{flushleft}


\section{Introduction} \label{sec:intro}

Failing to correctly identify the structure of the experimental design can lead to incorrect model selection and misleading statistical inferences. \citet{bate2016a} describes an approach to construct a list of \textit{structural objects}, known as the \textit{layout structure} (LS), which identifies the structure of the experimental design. The Hasse diagram is a useful way of visualising the layout structure as it describes the relationship between the experimental factors and the underlying structure of the experimental design. Including degrees of freedom on the diagram highlights any potential weaknesses the design may have. These are important precursors to the randomisation and subsequent statistical analysis.
 
The randomisation performed can influence which model terms, corresponding to structural objects in the layout structure, can be included in the statistical model. \citet{bate2016b} describes an approach that uses the layout structure, in conjunction with the randomisation, to construct a reduced list of objects, defined as \textit{randomisation objects}. This reduced list is known as the \textit{restricted layout structure} (RLS) and can be used to identify which terms to include in the statistical model. Randomisation objects, which form the restricted layout structure, can also be visualised using a generalisation of the Hasse diagram. This version of the Hasse diagram includes arrows, known as randomisation arrows, to illustrate the randomisation performed.

This paper describes the \texttt{hassediagrams} package in \texttt{R} \citep{R} that provides the user with a tool to generate Hasse diagrams of the layout and restricted layout structures, following the methodology developed in \citet{bate2016a} and \citet{bate2016b}. Although several \texttt{R} packages have been developed to address related tasks in the fields of experimental design, combinatorial mathematics and graphical models, the \texttt{hassediagrams} package extends these capabilities by offering the user the ability to generate the Hasse diagram of the layout and restricted layout structures. 

A closely related package is \texttt{dae} \citep{brien2024} which stands for Design and Analysis of Experiments. This package provides extensive tools for analysing data generated when using single-tier and multi-tier randomisation-based experimental designs. While \texttt{dae} focuses on the statistical aspects of understanding complex experimental designs, its capabilities do not cover the \texttt{hassediagrams} package's focus on visualising the structure of these designs. The \texttt{agricolae} package (\citep{de2021}, \citep{de2019}) is widely used for managing agricultural experimental designs and offers robust tools for generating and analysing treatment effects in designs such as Latin squares and randomised blocks. However, \texttt{agricolae}'s functionality does not support Hasse diagrams.

Beyond experimental designs, general-purpose graph visualisation packages such as \texttt{igraph} \citep{csardi2013} and \texttt{ggraph} \citep{pedersen2017} offer functionality similar to the \texttt{hassediagrams} package. The \texttt{igraph} package is a comprehensive procedure for constructing and analysing network structures, which share similarities with the hierarchical structures represented by Hasse diagrams. The \texttt{ggraph} package, built on the \texttt{ggplot2} framework, offers tailored tools for visualising hierarchical layouts and trees. 

Lastly, mathematical packages such as \texttt{partitions} and \texttt{relations} provide combinatorial and set-theoretical tools that underpin the structures visualised by Hasse diagrams.

In the context of experimental designs, \texttt{hassediagrams} is a powerful package that contributes to the statistical software ecosystem by providing specific tools for visualising the hierarchical relationships between factors as defined by the experimental design. The package also allows the user to visualise the randomisation performed, and identify the final statistical model. To the best of our knowledge, \texttt{hassediagrams} is the first \texttt{R} package that serves this purpose.

\section{Key concepts and notation} \label{sec:definitions}

Understanding the relationships between factors is crucial when identifying the structure of the experimental design. In general, factors can be related in four ways, \textit{fully crossed}, \textit{partially crossed}, \textit{nested} or \textit{equivalent}. For ease of defining key concepts and for introducing notation, consider two factors $A$ and $B$. 

Factor $A$ is said to be nested within factor $B$, denoted as $A(B)$, if each level of $A$ occurs with one and only one level of $B$ but at least one level of $B$ occurs with multiple levels of $A$ \citep[Chapter 7]{montgomery2017}. In this case, $A$ is said to be finer than $B$, and $B$ is coarser than $A$. 

The two factors, $A$ and $B$, are fully crossed if all levels of $A$ occur with all levels of $B$ and vice versa \citep[Chapter 7]{montgomery2017}. Following \citet{bailey1996} and \citet{tjur1984}, a generalised factor whose levels correspond to combinations of the crossed factors $A$ and $B$ is defined as:
\[
A \land B.
\]

Two factors, $A$ and $B$, are partially crossed if they are not fully crossed, but at least one level of $A$ occurs with more than one level of $B$ and vice versa.

Finally, two factors $A$ and $B$ are defined as equivalent if, for every occurrence of a level of $A$ within the design, the same level of $B$ occurs, and vice versa. In other words, the two factors are identical apart from the names of their levels, see \citet[p. 170]{bailey2008} and \citet{tjur1984}.

Once the experiment scheme has been developed and the experimental design selected, the researcher can perform the randomisation through the randomisation of the levels of one factor/generalised factor to another. This step is crucial for ensuring valid statistical inference and reflects the principles of randomisation outlined in \citet{preece1978}, which include: (i.) randomising the observational units, (ii.) randomising treatment labels, and (iii.) selecting a design randomly from a suitable set.

These concepts can be extended to facilitate randomisation-based models once the selected design has been randomised \citep{brien2006}. In \citet{bate2016b} the different types of randomisation are defined using randomisation arrows, where the objects at either end of an arrow are defined as \textit{randomisation objects}. There are many types of randomisation that can be performed:
\begin{enumerate} [label=\roman*)]
\item if the levels of factor $A$ are randomised to factor $B$, without involving any additional factor(s), then it is written as:
\[
A \to B,
\]
\item if the randomisation of $A$ to $B$ occurs separately at each level of a third factor $C$, then it is written as:
\[
A \to B[C],
\]
\item if factor $A$ is randomised by independently randomising factors $B$ and $C$, then the randomisation is written as:
\[
A \to B \otimes C,
\]
where the $\otimes$ symbol indicates separate permutation of levels for $B$ and $C$,
\item if the combinations of the levels of the crossed factors $A$ and $B$ are randomised to factor $C$, then this randomisation is expressed as:
\[
A \land B \to C,
\]
\item if the levels of generalised factor $A \land B$ are randomised to $C$ by independently randomising the levels of $A$ to $C$ and then $B$ to $C$, the notation is:
\[
A \otimes B \to C.
\]
\end{enumerate}

\section{Motivation example} \label{sec:toyexample}


As an example to motivate the use of Hasse diagrams, consider an experiment conducted to assess the benefit of using different types of polymer scaffold, seeded with neural stem cells, to aid in recovery following spinal cord injury (\citep{bate2014} and \citep{teng2002}). Three types of scaffold were assessed (implant only, neural stem cells only, implant and neural stem cells and a sham control). Forty-eight rats, twelve per treatment, were assigned to these treatment groups. As the surgery to implant the implants was conducted over three days, and it was felt this may impact the results, it was decided to use a block design (with Day as a blocking factor). Rats were assigned to days and then four rats per day randomly assigned to the implant treatment groups. 

\begin{figure}[H]
\centering
\begin{minipage}[t]{0.28\textwidth}
  \centering
  \includegraphics[width=\linewidth]{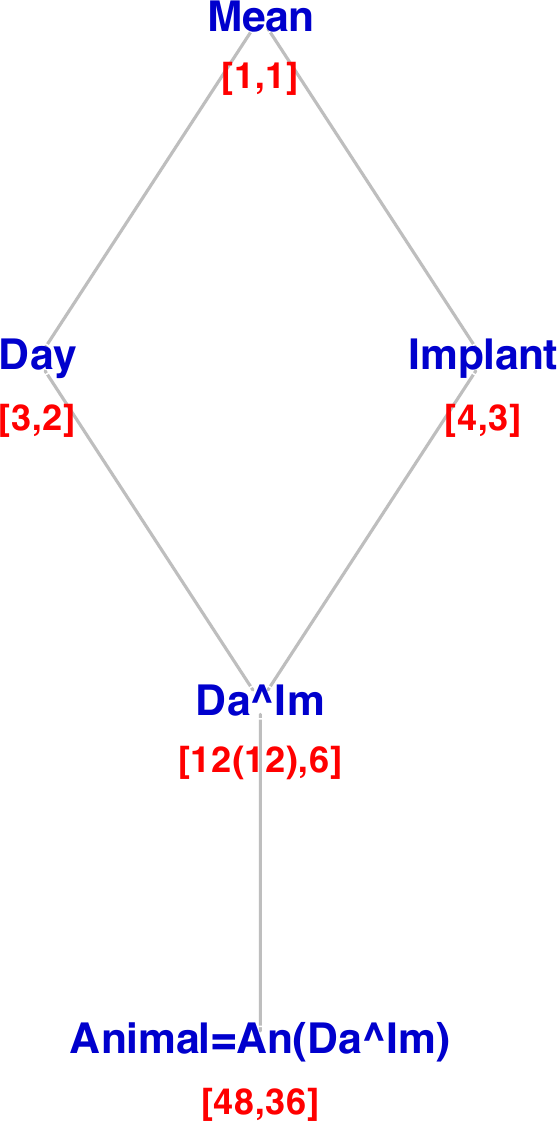}
\end{minipage}
\hspace{0.05\textwidth}
\begin{minipage}[t]{0.28\textwidth}
  \centering
  \includegraphics[width=\linewidth]{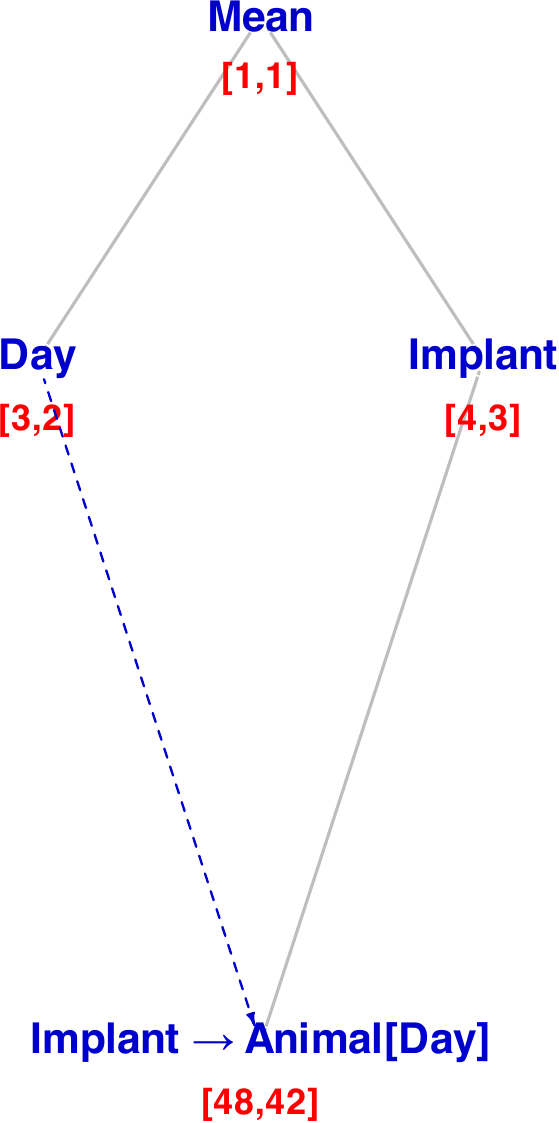}
\end{minipage}
\caption{(a) Hasse diagram of the layout structure of the motivation example, and (b) Hasse diagram of the restricted layout structure of the motivation example.}
\label{fig:toyexample}
\end{figure}

The layout structure represents how these experimental units (rats) and factors (Implant and Day) are organised prior to considering randomisation; see diagram (a) in Figure~\ref{fig:toyexample}. However, once randomisation is applied, where animals are randomly assigned to implants within each day, the restricted layout structure modifies the relationships; see diagram (b) in Figure~\ref{fig:toyexample}. Notation and implementation are described in Sections \ref{sec:layout} and \ref{sec:reslayout} with detailed examples in Section \ref{sec:examples}.

\section{Layout structure} \label{sec:layout}

The layout structure consists of a list of structural objects that correspond to the factors and generalised factors and a description of the relationships between them, as defined by the experimental design. It is constructed by considering the crossed and nested relationships between the factors within the experimental design. The layout structure can be illustrated using a generalisation of the Hasse diagram. The diagram consists of the structural objects in the layout structure, with objects corresponding to random effects underlined to differentiate them from objects corresponding to fixed effects. Additionally, the Hasse diagram can include (i) the number of levels of the structural object, for objects corresponding to factors, (ii) the maximum number of potential levels, for objects corresponding to generalised factors, (iii) the actual number of levels of the generalised factor included in the design, and (iv) the ANOVA-like degrees of freedom.

The procedure for identifying the structure of the experimental design consists of five stages, as described in \citet{bate2016a} and implemented in \texttt{hassediagrams}. The first stage involves listing all sources of variability that may influence the experimental response. This list of effects is informed by the hypotheses to be assessed and background knowledge of the experimental material. For each identified effect, a factor (that defines the experimental design) is generated. These include 'factors of interest', to assess the effects of interest, and 'nuisance or blocking factors', to allow additional sources of variability to be accounted for.

In the second stage, researchers must uniquely define the levels of each factor. This step is essential to algorithmically identify the structure of the design.

Following this, in stage three, the experiment scheme is generated. The experiment scheme describes the planning stages of the experiment and the decisions taken before randomising the experimental material. It consists of (i) a list of the factors and factor levels, as described in stages 1 and 2, (ii) a list of all the properties that the experimental design should have and (iii) the type of design required, (iv) the non-randomised design itself or, if unavailable, the randomised design instead.

Stage four focuses on summarising and visualising the relationships between these factors, using the layout structure. The layout structure consists of (i) a list of structural objects consisting of the factors that define the experimental design (identified in stage 1), (ii) the generalised factors that are implied by the structure of the experimental design, (iii) a description of the nested, crossed, and equivalent relationships between the factors, as defined by the experimental design. The layout structure is then visualised using a Hasse diagram.

Finally, in stage five, the degrees of freedom (DF) can be calculated by the subtraction method \citep{tjur1984}. This can be easily performed in conjunction with the Hasse diagram. This will identify any potential weaknesses in the design, for example caused by a lack of replication.

\subsection[Implementation of hasselayout()]{Implementation of \texttt{hasselayout}} \label{subsec:hasselayout}

The \texttt{hasselayout} function is designed to generate a Hasse diagram of the layout structure with the aim to assist users in visualising and interpreting the hierarchical structure of the experimental design, supporting better-informed decision-making during the analysis phase. This function takes a data set which contains the factors in the experimental design, and produces a graphical representation of the relationships between the aforementioned factors, i.e., whether they are fully crossed, partially crossed, nested, partially crossed, or equivalent. 

The function \texttt{hasselayout} is used as follows:
\begin{verbatim}
hasselayout(datadesign, randomfacsid = NULL, 
            showLS = "Y", showpartialLS = "Y", showdfLS = "Y",
            check.confound.df = "Y", maxlevels.df = "Y",
            table.out = "N", pdf = "N", example = "example", 
            outdir = NULL, hasse.font = "sans", produceBWPlot = "N", 
            structural.colour = "grey", structural.width = 2, 
            partial.colour = "orange", partial.width = 1.5,
            objects.colour = "mediumblue", df.colour = "red",
            larger.fontlabelmultiplier = 1, 
            middle.fontlabelmultiplier = 1, 
            smaller.fontlabelmultiplier = 1)
\end{verbatim}

The core and only mandatory input of the function is a data frame, passed to \texttt{hasselayout} via the \texttt{datadesign} argument, containing the factors that define the experimental design. More details of the mandatory \texttt{datadesign} argument is available in Table~\ref{table:hasselayoutMANDATORY}.

\begin{longtable}{p{5.5cm}p{9cm}}
\hline
 \textbf{Argument} & \textbf{Description}  \\
 \hline
\texttt{datadesign} & A data frame, list or environment (or object coercible by \texttt{as.data.frame} to a data frame) containing the variables/factors in the experimental design. The data frame should only include the variables/factors/columns that the user wants to include in the Hasse diagram. \\
\hline
\caption{The mandatory argument of the \texttt{hasselayout} function.}
\label{table:hasselayoutMANDATORY}
\end{longtable}

As well as the single mandatory argument, \texttt{hasselayout} holds several optional arguments all of which are detailed in Table~\ref{table:hasselayoutOPTIONAL}. The main optional argument is \texttt{randomfacsid}, used to indicate if factors are fixed or random. This argument must be a vector of length equal to the number of factors in the datadesign dataframe. The entries are either 0, for fixed factors, or 1 for random factors. The default option is all factors defined as fixed. 

The \texttt{hasselayout} function first identifies the relationships between the factors (e.g., fully crossed, partially crossed, nested or equivalent) by evaluating the levels in \texttt{datadesign}. To construct the Hasse diagram, the argument \texttt{showLS} must be set to its default option, i.e., set to "Y", otherwise the diagram is not generated. Suppressing the diagram generation may be useful for users who are interested in the tabulated output without needing a graphical representation. 

The \texttt{showpartialLS} argument controls the inclusion of dotted lines on the Hasse diagram that illustrate partially crossing between factors. When set to "Y", the default option, dotted lines are drawn between partially crossed factors, providing a clear visual indication of their relationship. If set to "N", these dotted lines will not be displayed, simplifying the diagram but potentially losing some detail of the factor relationships.

The \texttt{showdfLS} argument specifies whether to display the degree of freedom labels associated with each structural object on the Hasse diagram. By default, \texttt{showdfLS} is set to "Y", meaning that the labels will be shown. Setting \texttt{showdfLS} to "N" will suppress these labels, resulting in a cleaner diagram that focuses solely on the relationships among the factors without additional annotations. 

The \texttt{hasselayout} function can detect confounded degrees of freedom across the factors. This feature is controlled through the \texttt{check.confound.df} argument, where the default is "Y", i.e., a check for confounded degrees of freedom is performed. However, for larger experimental designs where checking for confounded degrees of freedom can be computationally intensive, setting \texttt{check.confound.df} to "N" allows users to bypass this step.

Additionally, \texttt{hasselayout} offers flexibility in controlling the Hasse diagram's output location, font, font sizes of the labels and the appearance of the plot. In context, the user can specify whether the Hasse diagram is output as a PDF file (\texttt{pdf} = "Y") or visualised directly in the \texttt{R} environment (\texttt{pdf} = "N"). If \texttt{pdf} = "Y", the user can control the name and location of the pdf file through arguments \texttt{example} and \texttt{outdir}, respectively. Control of the font and font sizes of the labels on the plot, including options for generating black-and-white diagrams, are available through the arguments \texttt{hasse.font}, \texttt{produceBWPlot}, \texttt{smaller.fontlabelmultiplier}, \texttt{middle.fontlabelmultiplier}, and \texttt{larger.fontlabelmultiplier}. The optional arguments \texttt{structural.width} and \texttt{partial.width} control the width of the structural object lines and the width of the partial crossing dotted lines, respectively. The plot colours can also be manipulated. The option \texttt{structural.colour} controls the colour of the structural lines, \texttt{partial.colour} controls the colour of the partial crossing dotted lines, \texttt{objects.colour} controls the structural object label font colour and \texttt{df.colour} controls the degrees of freedom font label colour.  
In addition to the graphical output, if the \texttt{table.out} option is enabled, i.e., \texttt{table.out} = "Y", then the function can generate the 'layout structure table', (see \citet{bate2016a}). The default setting is "N". This table provides a detailed account of the generalised factors, with entries indicating whether the factors are fully crossed (notated as 0), nested (notated as 1), or partially crossed (notated as (0)). In cases where the design involves confounded degrees of freedom, a table of the structural objects and their associated degrees of freedom is printed automatically.

\begin{longtable}{p{5.5cm}p{9cm}}
\hline
 \textbf{Argument} & \textbf{Description}  \\
 \hline
   \texttt{randomfacsid} & An optional vector specifying whether the factors are defined as fixed (entry = 0) or random (entry = 1). The default choice is NULL and the function automatically sets all entries to 0. The length of the vector should be equal to the number of variables/factors in the design, i.e., the length of the vector should be equal to the number of columns of the argument \texttt{datadesign}. \\
 \hline
 \texttt{showLS} & logical. If "N" then generation of the Hasse diagram is suppressed. The default is "Y". \\
  \hline
  \texttt{showpartialLS} & logical. If "N" then the partial crossing between structural objects (using dotted connecting lines) is not illustrated on the Hasse diagram of the layout structure. The default is "Y". \\
  \hline
  \texttt{showdfLS} & logical. If "N" then the structural object label is not displayed on the Hasse diagram of the layout structure. The default is "Y". \\
  \hline
  \texttt{check.confound.df} & logical. If "N" then the check for confounded degrees of freedom is not performed. The default is "Y". \\
  \hline
  \texttt{maxlevels.df} & logical. If "N" then the potential maximum number of levels of a generalised factor is removed from the structural object label on the Hasse diagram of the layout structure. The default is "Y". \\
  \hline
  \texttt{table.out} & logical. If "Y" then a table that shows the relationships between the structural objects in the layout structure is printed. The default is "N". \\
  \hline
  \texttt{pdf} & logical. If "Y" then a pdf file containing the Hasse diagram of the layout structure is generated. The default is "N", i.e., a pdf file is not generated. \\
  \hline
  \texttt{example} & File name for the pdf output file containing the Hasse diagram. The default is set to "example". \\
  \hline
  \texttt{outdir} & Location of the pdf output file if \texttt{pdf}="Y". The default is set to NULL and in this case the pdf output file containing the Hasse diagram output is stored to a temporary file. To specify a permanent location this argument needs be specified. \\
  \hline
  \texttt{hasse.font} & The name of the font family used for all text included in the Hasse diagram. Standard and safe font families to choose are "sans", "serif", and "mono". If the design's factor labels contain Unicode characters, or for consistency with Hasse diagrams of restricted layout structures using \texttt{hasserls}, a Unicode friendly font family should be selected. More details on Unicode friendly family options are in Section \ref{subsec:hasserls}. The default is "sans". \\
  \hline
  \texttt{produceBWPlot} & logical. If "Y" then the Hasse diagram will be generated in black and white format. The default is set to "N", i.e., a coloured version of the plot is produced. \\
  \hline
  \texttt{structural.colour} & The colour of the structural lines that connect structural objects on the Hasse diagram. The default colour is "grey". \\
  \hline
  \texttt{structural.width} & The width of the structural lines on the Hasse diagram. The default width is 2. \\
  \hline
  \texttt{partial.colour} & The colour of the dotted lines on the Hasse diagram connecting partially crossed objects. The default colour is "orange". \\
  \hline
  \texttt{partial.width} & The width of the partial crossing dotted lines on the Hasse diagram. The default width is 1.5. \\
  \hline
  \texttt{objects.colour} & The colour of the labels of the structural objects on the Hasse diagram. The default colour is "mediumblue". \\
  \hline
  \texttt{df.colour} & The colour of the degrees of the freedom labels on the Hasse diagram. The default colour is "red". \\
  \hline
  \texttt{larger.fontlabelmultiplier} & The large font multiplier is the multiplier for the font used for the labels of objects on the Hasse diagram where there are four or less objects at that level in the diagram. The default is 1. \\
  \hline
  \texttt{middle.fontlabelmultiplier} & The medium font multiplier is the multiplier for the font used for the labels of objects on the Hasse diagram involving a factor that is equivalent to a generalised factor. The default is 1. \\
  \hline
  \texttt{smaller.fontlabelmultiplier} & The small font multiplier is the multiplier for the font used for the labels of objects on the Hasse diagram where there are five or more objects at that level of the diagram. The default is 1. \\
  \hline
\caption{The optional arguments of the function \texttt{hasselayout}.}
\label{table:hasselayoutOPTIONAL}
\end{longtable}

\section{Restricted layout structure} \label{sec:reslayout}

Once the structure of the experimental design has been defined and the layout structure generated, the randomisation is then performed. 

As discussed above, the individual randomisations can be described using randomisation arrows. Objects at either end of these arrows (which correspond to structural objects in the layout structure) are part of the set of objects known as randomisation objects. Randomisation arrows are oriented to either (i) start from factors higher up in the Hasse diagram and point to factors below them or (ii) involve equivalent objects within a single label on the Hasse diagram. This approach is discussed in \citet{bate2016b} and is an adaptation of an earlier methodology in \citet{brien2006}. It enables a clear distinction between randomisation objects (that are implied by the randomisation) and structural objects (that define the relationships between factors in the design).

The \texttt{hassediagrams} package uses the notation from Section \ref{sec:definitions} to define the randomisation consistently across designs. Each randomisation object corresponds to a structural object, and their levels are aligned. For example, the randomisation object $B \otimes C$ corresponds to the structural object $B \land C$, and their levels represent combinations of the levels of factors $B$ and $C$. The corresponding structural objects, randomisation objects and hence the terms in the statistical model are given in Table~\ref{table:rlsobjects}.

\begin{table}[ht]
\centering
\begin{tabular}{|l|l|l|}
\hline
\textbf{Structural object} & \textbf{Randomisation object} & \textbf{Model term} \\
\hline
$A$ & $A$ & $A$ \\
$A \land B$ & $A \land B$ & $A * B$ \\
$A(B)$ & $A[B]$ & $A * B$ \\
$A \land B$ & $A \otimes B$ & $A * B$ \\
$A \land B(C)$ & $\{A \otimes B\}[C]$ & $A * B * C$ \\
$A \land B(C)$ & $A \otimes \{B[C]\}$ & $A * B * C$ \\
$A(B \land C)$ & $A[B \otimes C]$ & $A * B * C$ \\
$A \land B \land C$ & $A \otimes B \otimes C$ & $A * B * C$ \\
\hline
\end{tabular}
\caption{List of corresponding structural objects, randomisation objects and model terms.}
\label{table:rlsobjects}
\end{table}

Once the randomisation has been defined the next step is to construct the restricted layout structure. The restricted layout structure consists of a list of randomisation objects (that correspond to structural objects in the layout structure) and a description of the relationships between the randomisation objects, defined as randomisation-nesting, see Table~\ref{table:randomisation}. 

\begin{table}[ht]
\centering
\begin{tabular}{|l|p{6.5cm}|}
\hline
\textbf{Randomisation object} & \textbf{Randomisation-nest objects} \\
\hline
$A$ & Mean \\
$A \land B$ & Mean, $A$, $B$ \\
$A[B]$ & Mean, $B$ \\
$A \otimes B$ & Mean, $A$, $B$ \\
$\{A \otimes B\}[C]$ & Mean, $A[C]$, $B[C]$, $C$ \\
$A \otimes \{B[C]\}$ & Mean, $A \otimes C$, $B[C]$, $A$, $C$ \\
$A[B \otimes C]$ & Mean, $B \otimes C$, $B$, $C$ \\
$A \otimes B \otimes C$ & Mean, $A$, $B$, $C$, $A \otimes B$, $A \otimes C$, $B \otimes C$ \\
\hline
\end{tabular}
\caption{Randomisation objects and the objects that randomise-nest them.}
\label{table:randomisation}
\end{table}

The restricted layout structure can now be created using a set of four rules, as defined in \citet{bate2016b}.

\textit{Rule 1}: To begin, the randomisation objects that are at the beginning and end of a randomisation arrow are included in the restricted layout structure.

\textit{Rule 2}: All randomisation objects that randomisation-nest the randomisation objects in Rule 1 are included in the restricted layout structure.

\textit{Rule 3}: The randomisation object that corresponds to the mean and the randomisation object that corresponds to the generalised factor that defines the observational unit are included, regardless of whether the latter is involved in the randomisation or not.

\textit{Rule 4}: All randomisation objects that correspond to a fixed generalised factor of the form $(F_{1} \land F_{2} \land \dots \land F_{p})$, with $F_{1}, F_{2}, \dots, F_{p}$ being factors, are included in the restricted layout structure if (i) the corresponding generalised factor exists in the layout structure and (ii) the fixed factors that define it, i.e., $F_{1}, F_{2}, \dots, F_{p}$, are themselves randomised to other factors or generalised factors.

This list is therefore constructed by considering the randomisation in conjunction with the layout structure and whether the factors are defined as fixed or random. This structure can be visualised using a Hasse diagram.

\subsection[Implementation of itemlist()]{Implementation of \texttt{itemlist}} \label{subsec:itemlist}

The \texttt{itemlist} function is a key component in structuring the factors of an experimental design for the purpose of generating a Hasse diagram of the restricted layout structure. It returns an object of class \texttt{rls}, which contains the structural objects from the layout structure of the design and their relationships in a form that can be used for further analysis. This structured output is specifically formatted to be utilised by the \texttt{hasserls} function to visualise the relationships between the randomisation objects. Importantly, \texttt{itemlist} serves as a mandatory intermediate step before applying the \texttt{hasserls} function.

The function \texttt{itemlist} is used as follows:
\begin{verbatim}
itemlist(datadesign, randomfacsid=NULL)
\end{verbatim}

The \texttt{itemlist} function consists of two arguments; \texttt{datadesign} which is a mandatory input, and \texttt{randomfacsid} which is an optional input that defaults to NULL. Since both arguments were thoroughly explained in the context of the \texttt{hasselayout} function earlier in this paper, specifically in Section \ref{subsec:hasselayout}, they will be briefly summarised here. A detailed documentation of \texttt{datadesign} is available in Table~\ref{table:hasselayoutMANDATORY}, and a detailed documentation of \texttt{randomfacsid} is available in Table~\ref{table:hasselayoutOPTIONAL}.

As with \texttt{hasselayout}, \texttt{datadesign} is a data frame that contains the experimental design factors. These factors define the layout structure and the relationships between them, such as nesting or crossing. It is essential that the user provides only the columns relevant to the factors of the design under investigation (i.e., no response variable(s)).

The optional argument \texttt{randomfacsid} is a vector that allows the user to specify which factors are fixed (with a value of 0) and which are random (with a value of 1). If this argument is not provided, all factors are defined as fixed by default. As with \texttt{datadesign}, this input mirrors the structure used in \texttt{hasselayout}.

The \texttt{itemlist} function generates an object of class \texttt{rls}, which is a list containing multiple components that define the layout structure. The \texttt{rls} class object holds key information such as: the names of the structural objects (factors and generalised factors) present in the layout structure, the relationships between factors, and a matrix named \texttt{TransferObject} used for specifying randomisation objects to pass to \texttt{hasserls}.

To assist in presenting the important output returned by \texttt{itemlist}, custom print and summary methods have been defined for objects of class \texttt{rls}. The print/summary option is designed to streamline the interpretation of the output by highlighting key components necessary for generating the Hasse diagram of the restricted layout structure. The \texttt{print.rls} and \texttt{summary.rls} functions display a matrix of two columns. The first column contains the names of all the structural objects in the layout structure and the second column contains the Mean and NULL values that need to be filled in to define the corresponding randomisation objects. The aim of this output is to provide a concise view of the structural objects, enabling the user to quickly identify the necessary input for the next step in the analysis, using the \texttt{hasserls} function.

\subsection[Implementation of hasserls()]{Implementation of \texttt{hasserls}} \label{subsec:hasserls}

The \texttt{hasserls} function is designed to generate a Hasse diagram of the restricted layout structure. This function requires an object of class \texttt{rls}, which can be generated using the \texttt{itemlist} function, that contains the necessary structural information of the experimental design.

The function \texttt{hasserls} is used as follows:
\begin{verbatim}
hasserls(object, randomisation.objects, random.arrows = NULL, 
           showRLS = "Y", showpartialRLS = "Y", showdfRLS = "Y", 
           showrandRLS = "Y", check.confound.df = "Y", 
           maxlevels.df = "Y", table.out = "N", equation.out = "N",
           pdf = "N", example = "example", outdir = NULL, 
           hasse.font = "sans", produceBWPlot = "N", 
           structural.colour = "grey", structural.width = 2,
           partial.colour = "orange", partial.width = 1.5,
           objects.colour = "mediumblue", df.colour = "red",
           arrow.colour = "mediumblue", arrow.width = 1.5,
           arrow.pos = 7.5, larger.fontlabelmultiplier = 1, 
           middle.fontlabelmultiplier = 1, 
           smaller.fontlabelmultiplier = 1)
\end{verbatim}

The function consists of two mandatory arguments, both of which are explained in Table~\ref{table:hasserlsMANDATORY}. The first mandatory argument is \texttt{object} which is a class \texttt{rls} object containing structural objects of the layout structure (generated using the \texttt{itemlist} function). The second mandatory argument is \texttt{randomisation.objects} which is a two-column matrix that includes structural objects in the first column and user-defined randomisation objects in the second column. The randomisation objects should correspond to structural objects and must be manually defined after creating the \texttt{rls} object. The \texttt{randomisation.objects} argument is a user modified version of the \texttt{TransferObject} item from the \texttt{itemlist} output. The user is therefore required to modify the second column of the 'randomisation objects' matrix to specify the randomisation objects that are part of the restricted layout structure. Any structural objects not included in the restricted layout structure should be left as "NULL" in the second column of the 'randomisation objects' matrix. The edited entries in the second column correspond to the labels in the randomisation objects as they appear on the Hasse diagram of the restricted layout structure. 

\begin{longtable}{p{5.5cm}p{9cm}}
\hline
 \textbf{Argument} & \textbf{Description}  \\
 \hline
\texttt{object} & An object of class \texttt{rls}. The function \texttt{itemlist} generates the class \texttt{rls} object. Printing the \texttt{rls} object will give a list of structural objects (that define the layout structure) to aid in defining the randomisation objects in the restricted layout structure. \\
\hline
\texttt{randomisation.objects} & This argument takes the format of the \texttt{TransferObject} item from the class \texttt{rls} object. The first column contains the names of all the structural objects in the layout structure (automatically generated) and the second column contains the corresponding randomisation objects in the restricted layout structure (manually generated). To begin with, the function \texttt{itemlist} should be ran to generate the class \texttt{rls} object. The user will then need to edit the second column of the \texttt{TransferObject} matrix to define the randomisation objects that appear in the restricted layout structure. Structural objects that do not occur in the restricted layout structure must be left as "NULL" in the second column. The names specified in the second column represent the labels of the randomisation objects on the Hasse diagram of the restricted layout structure. \\
\hline
\caption{The mandatory arguments of the \texttt{hasserls} function.}
\label{table:hasserlsMANDATORY}
\end{longtable}

Of the optional arguments, all of which are explained in detail in Table~\ref{table:hasserlsOPTIONAL}, the most important is \texttt{random.arrows}. This optional argument can be utilised to specify the randomisations performed using randomisation arrows. The default option is NULL, i.e., no randomisation arrows. To add randomisation arrows on the Hasse diagram of the restricted layout structure, \texttt{random.arrows} must be set as a two-column matrix of integers where each row defines a randomisation arrow. The entries in the first column contain the object(s) at the start of the randomisation arrow and the second column contains the object(s) at the end (where the values correspond to the entry number for the randomisation object in the \texttt{TransferObject} item from the class \texttt{rls} object). As discussed above arrows between non-equivalent objects should point downwards on the Hasse diagram and hence it is required that the second column entries are numerically larger than the first column entries. The argument \texttt{showrandRLS} manages the depiction of randomisation arrows on the diagram (default = "Y"). If \texttt{random.arrows} is NULL, this will automatically revert to "N". The position of the randomisation arrows on the Hasse diagram is adjusted by the \texttt{arrow.pos} argument. It specifies the distance between the arrows and the objects they point at; smaller values generate longer arrows, while larger values result in shorter arrows, with a default value of 7.5.

The remaining optional arguments provide flexibility for the user to change the formatting options and incorporate additional information on the Hasse diagram of the restricted layout structure. For instance, \texttt{pdf}, \texttt{outdir} and \texttt{example} (as described in Section \ref{subsec:hasselayout}) can be used to store the Hasse diagram in a PDF file of user-defined name and directory. Optional arguments that are used for additional information on the Hasse diagram and output, including \texttt{check.confound.df}, and \texttt{maxlevels.df} are also the same as described in Section \ref{subsec:hasselayout}. Arguments \texttt{showRLS}, \texttt{showpartialRLS}, and \texttt{showdfRLS} of \texttt{hasserls} are equivalent to \texttt{showLS}, \texttt{showpartialLS}, and \texttt{showdfLS} of \texttt{hasselayout}.

The user can manipulate the visualisation of the Hasse diagram, as described in Section \ref{subsec:hasselayout}. Along with the options described in Section \ref{subsec:hasselayout}, the optional argument \texttt{arrow.width} controls the width of the randomisation arrow(s) and \texttt{arrow.colour} controls the colour of the arrow(s).   

In addition to the graphical output, if the \texttt{table.out} option is enabled, i.e., \texttt{table.out} = "Y", the \texttt{hasserls} function generates a table showing the relationships between the randomisation objects in the restricted layout structure. The default is "N". Finally, an equation outlining a recommended mixed model to use in the stats analysis can be provided if \texttt{equation.out} = "Y", with the default being "N". 

In contrast to the layout structure, the restricted layout structure uses the symbols $\otimes$ and $\to$, as defined in Section~\ref{sec:definitions}, to represent the randomisation performed. In \texttt{R}, these symbols correspond to the Unicode characters \verb|\u2297| and \verb|\u2192|, respectively. Rendering these Unicode characters correctly in Hasse diagrams requires using a Unicode-friendly font, which can be non-trivial due to differences in font availability and rendering behaviour across platforms. For instance, Windows users may have access to fonts such as Cambria, Segoe UI Symbol, Ebrima, or Arial Unicode MS, while macOS users might rely on system fonts like AppleMyungjo, .SF Compact, or .SF NS Rounded. Moreover, not all installed fonts are automatically available to R's plotting devices.

To identify fonts that are available in an \texttt{R} session, users can run:
\begin{verbatim}
systemfonts::system_fonts()$family
\end{verbatim}

Fonts that are installed on the system, but not yet available to \texttt{R}, can be imported using either:
\begin{verbatim}
showtext::font_import()
\end{verbatim}
or
\begin{verbatim}
extrafont::font_import()
\end{verbatim}
depending on the package used for font management. The list of successfully imported fonts can then be retrieved with either:
\begin{verbatim}
showtext::fonts()
\end{verbatim}
or
\begin{verbatim}
extrafont::fonts()
\end{verbatim}

Users should note that some Unicode-supporting fonts, such as Arial Unicode MS, may need to be downloaded manually from external sources, as they are not pre-installed on all systems. Given these nuances, robust cross-platform Unicode rendering requires careful selection and validation of font families within the Hasse diagram rendering function.

To address font compatibility issues, users may set the argument \texttt{pdf = "Y"}, which renders the Hasse diagram using the \texttt{cairo\_pdf()} device and saves it as a PDF file. This approach provides more robust Unicode support, and the default \texttt{sans} font is typically sufficient for rendering $\otimes$ and $\to$ correctly. The resulting PDF file is named using the value provided in the \texttt{example} argument (default: \texttt{"example"}). By default, the output is stored in a temporary location, but a permanent location can be specified using the \texttt{outdir} argument. In the examples section of this paper, where Unicode characters are used in the Hasse diagrams of the restricted layout structures, the \texttt{pdf = "Y"} option is enabled to ensure correct rendering.

\begin{longtable}{p{5.5cm}p{9cm}}
\hline
 \textbf{Argument} & \textbf{Description}  \\
 \hline
   \texttt{random.arrows} & A matrix of two columns that takes integer entries. The entries in the first column define the object(s) at the start of the randomisation arrow and the second column define the object(s) at the end. The values correspond to the entry number for the randomisation object in the \texttt{TransferObject} item from the class \texttt{rls} object. The first column specifies the randomisation object(s) at the beginning of the randomisation arrow(s) and the second column specifies the randomisation object(s) at the end of the arrow. The randomisation arrows must point downwards, hence, in each row of the matrix the second column entry must be larger than the first column entry. The randomisation object(s) involved in the randomisation arrow(s) must first be specified in the randomisation.objects argument. \\
   \hline
   \texttt{showRLS} & logical. If "N" then generation of the Hasse diagram of the restricted layout structure is suppressed. The default is "Y". \\
  \hline
  \texttt{showpartialRLS} & logical. If "N" then the partial crossing between randomisation objects (using dotted connecting lines) is not included  on the Hasse diagram of the restricted layout structure. The default is "Y". \\
  \hline
  \texttt{showdfRLS} & logical. If "N" then the randomisation object label is not displayed on the Hasse diagram of the restricted layout structure. The default is "Y". \\
  \hline
  \texttt{showrandRLS} & logical. If "N" then the randomisations are not illustrated (using arrows) on the Hasse diagram of the restricted layout structure. The default is "Y". If random.arrows=NULL, then \texttt{showrandRLS} defaults to "N". \\
  \hline
  \texttt{check.confound.df} & logical. If "N" then the check for confounded degrees of freedom is not performed. The default is "Y". \\
  \hline
  \texttt{maxlevels.df} & logical. If "N" then the potential maximum number of levels of a generalised factor is removed from the structural object label on the Hasse diagram of the restricted layout structure. The default is "Y". \\
  \hline
  \texttt{table.out} & logical. If "Y" then a table that shows the relationships between the randomisation objects in the restricted layout structure is printed. The default is "N". \\
  \hline
  \texttt{equation.out} & logical. If "Y" then a recommended mixed model to use in the statistical analysis is printed. The default is "N". \\
  \hline 
  \texttt{pdf} & logical. If "Y" then a pdf file containing the Hasse diagram of the restricted layout structure is generated. The default is "N", i.e., a pdf file is not generated. \\
  \hline
  \texttt{example} & File name for the pdf output file containing the Hasse diagram. The default is set to "example". \\
  \hline
  \texttt{outdir} & Location of the pdf output file if \texttt{pdf}="Y". The default is set to NULL and in this case the pdf output file containing the Hasse diagram output is stored to a temporary file. To specify a permanent location this argument needs be specified. \\
  \hline
  \texttt{hasse.font} & The name of the font family used for all text included on the Hasse diagram. Standard and safe font families to choose are "sans", "serif", and "mono". If any of the labels of the randomisation objects (as defined in the second column of \texttt{randomisation.objects} matrix) contain Unicode characters, a Unicode friendly font family should be selected. For more details on Unicode friendly family options see the Details section of the documentation of \texttt{hasserls}. If the font family selected fails to render, the font is automatically changed to "sans" instead. The default is "sans". \\
  \hline
  \texttt{produceBWPlot} & logical. If "Y" then the Hasse diagram will be generated in black and white format. The default is set to "N", i.e., a coloured version of the plot is produced. \\
  \hline
  \texttt{structural.colour} & The colour of the structural lines that connect structural objects on the Hasse diagram. The default colour is "grey". \\
  \hline
  \texttt{structural.width} & The width of the structural lines on the Hasse diagram. The default width is 2. \\
  \hline
  \texttt{partial.colour} & The colour of the partial crossing dotted lines of the connecting objects on the Hasse diagram. The default colour is "orange". \\
  \hline
  \texttt{partial.width} & The width of the partial crossing dotted lines on the Hasse diagram. The default width is 1.5. \\
  \hline
  \texttt{objects.colour} & The colour of the labels of the structural objects on the Hasse diagram. The default colour is "mediumblue". \\
  \hline
  \texttt{df.colour} & The colour of the degrees of the freedom labels on the Hasse diagram. The default colour is "red". \\
  \hline
  \texttt{arrow.colour} & The colour of the randomisation arrows on the Hasse diagram. The default colour is "mediumblue". \\
  \hline
  \texttt{arrow.width} & The randomisation arrows width on the Hasse diagram. The default width is 1.5. \\
  \hline
  \texttt{arrow.pos} & Specifies the position of the randomisation arrows, i.e., how far the randomisation arrows will be from the objects they point at. The default is 7.5. A smaller number specifies longer arrows and a higher number specifies shorter arrows. \\
 \hline
  \texttt{larger.fontlabelmultiplier} & The large font multiplier is the multiplier for the font used for the labels of objects on the Hasse diagram where there are four or less objects at that level in the diagram. The default is 1. \\
  \hline
  \texttt{middle.fontlabelmultiplier} & The medium font multiplier is the multiplier for the font used for the labels of objects on the Hasse diagram involving a factor that is equivalent to a generalised factor. The default is 1. \\
  \hline
  \texttt{smaller.fontlabelmultiplier} & The small font multiplier is the multiplier for the font used for the labels of objects on the Hasse diagram where there are five or more objects at that level of the diagram. The default is 1. \\
  \hline
\caption{The optional arguments of the function \texttt{hasserls}.}
\label{table:hasserlsOPTIONAL}
\end{longtable}

\section{Datasets} \label{sec:datasets}

The \texttt{hassediagrams} package includes four datasets designed to illustrate the functionality of the functions implemented within \texttt{hassediagrams}. Each of the datasets serves a specific purpose, providing users with examples of various types of experimental designs to enhance their understanding of the package's functionality. In this section, each dataset is described, including its source, structure, and factor definitions.

\subsection{Concrete dataset} \label{subsec:concrete}

The \texttt{concrete} dataset, involving the application of a fractional factorial design, was conducted to investigate the production of asphalt concrete \citep{anderson1974}.

This dataset consists of 16 experimental runs and six controllable factors, each of which are set at one of two levels per run. The categorical factors are Aggregate gradation (levels: 'fine' and 'course') and Curing condition (levels: 'wrapped' and 'unwrapped')and the continuous factors are Compaction temperature (levels: 250 and 300),  Asphalt content (levels: 5 and 7) and Curing temperature (levels: 45 and 72). Additionally, the run number is represented in the dataset as a categorical factor (levels: 1 to 16).

\subsection{Dental dataset} \label{subsec:dental}

The \texttt{dental} dataset contains the data from a crossover study aimed at investigating the effects of two chlorhexidine (CHX) rinses vs. saline on plaque regrowth over a period of four days. This design, sourced from \citet{newcombe1995}, involves 24 patients assessed across three treatment periods. The design is based on four pairs of 3 $\times$ 3 Latin squares, balanced to account for carry-over effects.

This dataset consists of 72 rows and five categorical factors: Sequence (levels: 1 to 6), Subject (levels: 1 to 32), Period (levels: 1 to 3), Treatment (levels: 'CHX1', 'CHX2', and 'saline'), and Observation (levels: 1 to 72).

\subsection{Human dataset} \label{subsec:human}

The \texttt{human} dataset contains the data of a block design used to compare two methods (mouse and stylus) for drawing a map in a computer file. This design involves 12 subjects randomised over six days, with two tests conducted within each day (morning and afternoon) across two different rooms. The design is based on 2 $\times$ 2 Latin squares, as detailed in \citet{brien2006}. 

This dataset consists of 24 rows and seven categorical factors. The factors include Subject (levels: 1 to 12), Day (levels: 1 to 6), Room (levels: 'A' and 'B'), Period (levels: 'morning' and 'afternoon'), Method (levels: 'mouse' and 'stylus'), Sequence (levels: 1 and 2) and Test (levels: 1 to 24).

\subsection{Analytical dataset} \label{subsec:analytical}

The \texttt{analytical} dataset is a cross-nested design used to assess the reliability of an analytical method, as described in \citet{bate2016a} and \citet{bate2016b}. This dataset is derived from an experiment involving three batches of material, analysed by four analysts, with two analysts working at each of two sites. Within each site, there are two chromatographic systems and two columns. For each combination of batch, analyst, system, and column, two preparations were made, and two injections were performed from each preparation. 

This dataset consists of 192 rows and eight categorical factors. The factors include Site (levels: 1 and 2), Analyst (levels: 1 to 4), Run  (levels: 1 to 16), Preparation (levels: 1 to 96), Injection (levels: 1 to 192), System (levels: 1 to 4), Column (levels: 1 to 4) and Batch (levels: 1 to 3).

\section{Examples} \label{sec:examples}

In this section, several examples are presented to demonstrate the functionality and versatility of the three core functions of the \texttt{hassediagrams} \texttt{R} package: \texttt{hasselayout}, \texttt{itemlist}, and \texttt{hasserls}. These examples illustrate the usage of the functions across different types of experimental designs, including factorial designs, balanced incomplete block designs (BIBD), crossover designs, and split-plot designs. Each example showcases how the functions can be applied to different experimental structures, highlighting the adaptability of the package to diverse applications and research areas. The examples include scenarios where specific settings are enabled or disabled, altering the output and the appearance of the Hasse diagrams. They therefore provide practical insights into the potential applications and flexibility of the \texttt{hassediagrams} package.

\subsection{Example. Balanced incomplete block design} \label{subsec:bibdexample}

\citet[Chapter 12]{joshi1987} describes an agricultural trial, based on a balanced incomplete block design, to investigate the yield of six varieties of wheat. The wheat varieties were allocated to thirty plots of land, where the land was separated into ten blocks with three plots in each block. The three categorical factors included in the experimental design are Variety (levels: 1 to 6), Block (levels: 1 to 10) and Plot (levels: 1 to 30). Within the experiment scheme a balanced incomplete block design was selected. The design and response is available in the \texttt{BIBDWheat.dat} dataset of the \texttt{dae} package in \texttt{R}. The dataset is loaded via the following code:

\begin{verbatim}
data(BIBDWheat.dat)
\end{verbatim}

Since only the design is required, it is very important to remove the response variable from the dataset before passing it to the functions in \texttt{hassediagrams}. The response variable, which is in the 4th column, is removed via:

\begin{verbatim}
BIBDWheat <- BIBDWheat.dat[ , -4]
\end{verbatim}

Note that, the levels of the factors must be uniquely identified (i.e., have a physical meaning) otherwise the function will not correctly identify the nesting and/or crossing of the factors. For this reason, an additional column Plots (levels: 1 to 30) is added to the dataset to index the rows of the design.

\begin{verbatim}
BIBDWheat$Plots <- c(1:30)
\end{verbatim}

This section will include Hasse diagrams of the layout structure and the restricted layout structure to demonstrate the full range of features provided by the functions in the \texttt{hassediagrams} package. Key features that are highlighted include:

\begin{itemize}
\item The generation of the Hasse diagrams of the layout and restricted layout structures (additional output options are discussed in later examples), hence the arguments \texttt{showLS} and \texttt{showRLS} are set to "Y" (the default options).
\item Inclusion of the degrees of freedom on the Hasse diagrams is achieved by setting the arguments \texttt{showdfLS} and \texttt{showdfRLS} to "Y" (the default options).
\item Partial crossing between the structural and randomisation objects is demonstrated on the Hasse diagrams of the layout and restricted layout structures by setting the arguments \texttt{showpartialRLS} and \texttt{showpartialRLS} to "Y", respectively (the default options).
\item The potential maximum number of levels of the generalised factors are included by setting the argument \texttt{maxlevels.df} being equal to "Y" (the default option).
\item A check for confounded degrees of freedom across factors is performed via setting the argument \texttt{check.confound.df} to "Y" (the default option).
\item The illustration of the randomisation (for the Hasse diagram of the restricted layout structure only) is achieved by setting the argument \texttt{showrandRLS} to "Y" (the default option).
\end{itemize}

All three factors are assumed to be fixed and this will be reflected in the code through the \texttt{randomfacsid} argument. This is a vector of length three, with all entries set to 0, i.e., fixed. The default option, which is NULL, is equivalent to all factors considered as fixed. 

To enhance readability, coloured versions of the Hasse diagrams will be generated (by leaving \texttt{produceBWPlot} to its default "N" option), with the font and small and medium font sizes set to their default options. The large font multiplier, which is applied when there are four or fewer objects at a particular level in the diagram, will be set to 1.8 to ensure clearer visibility of the objects. This adjustment makes interpretation of the Hasse diagrams easier.

\begin{figure}[H]
\centering
\includegraphics[width=\linewidth]{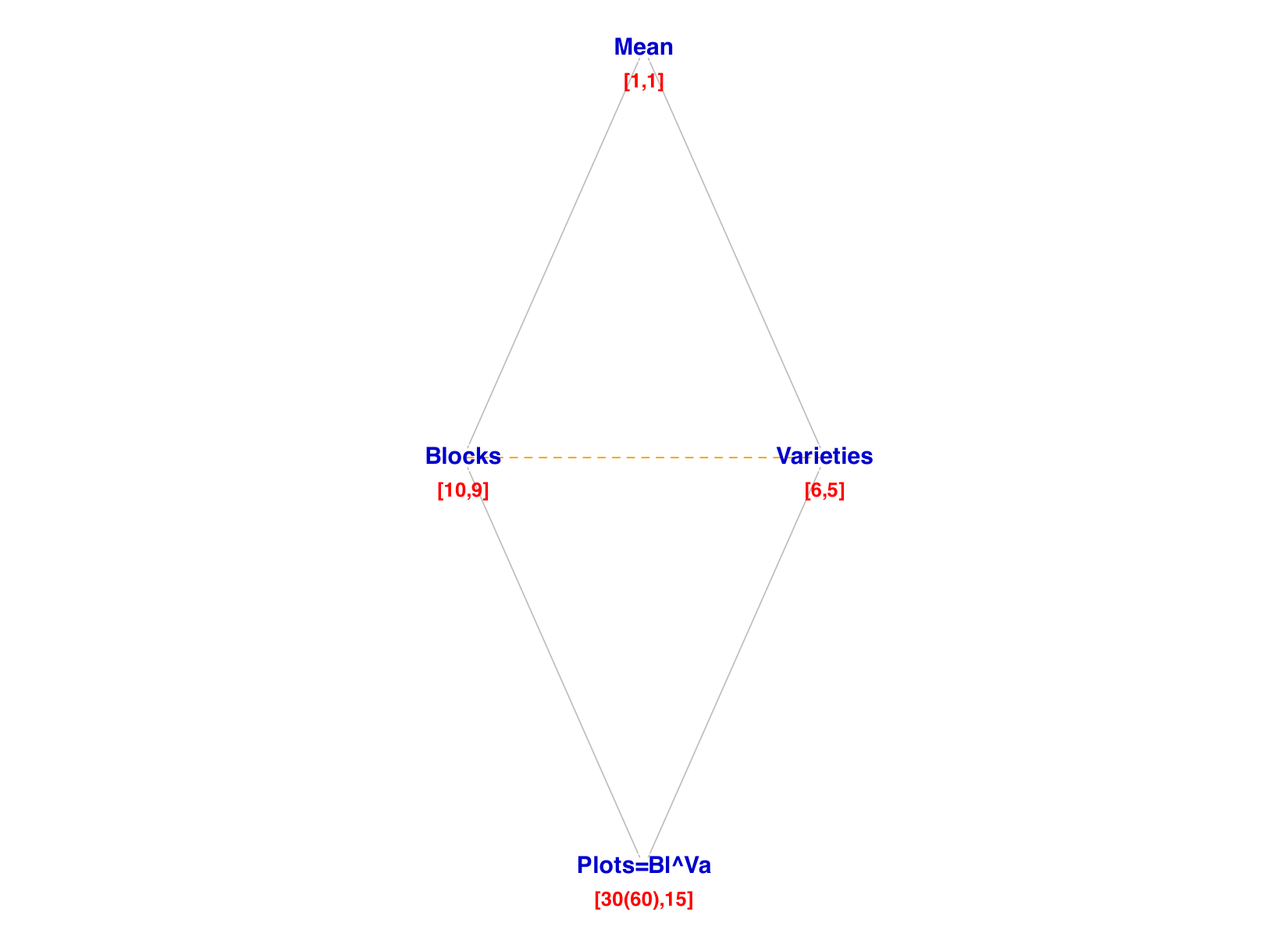}
\caption{Hasse diagram of the layout structure for Example \ref{subsec:bibdexample}}
\label{fig:bibdlayout}
\end{figure}

Starting with the layout structure, the Hasse diagram is generated using the \texttt{hasselayout} function:

\begin{verbatim}
hasselayout(datadesign = BIBDWheat, randomfacsid = c(0, 0, 0),
            showLS = "Y", showpartialLS = "Y",  showdfLS = "Y",
            check.confound.df = "Y", maxlevels.df = "Y",
            larger.fontlabelmultiplier = 1.8)
\end{verbatim}

The Hasse diagram of the layout structure is given in Figure~\ref{fig:bibdlayout}.

The experimental scheme included the decision on using the BIBD and also the pre-randomised allocation of varieties to blocks, based on the choice of balanced incomplete block. The randomisation performed was a conventional randomisation of a block design where the varieties allocated to each block randomised to plots separately for each block. The randomisation can be defined as: 

\begin{equation}
\label{eq:runrandom}
Variety \to Plot[Block]
\end{equation}

By following the rules described in Section \ref{sec:reslayout}, the randomisation objects that correspond to the structural objects $Blocks$, $Varieties$ and $Blocks \land Varieties$ will be included in the restricted layout structure, where the latter will be defined as $Plot[Block]$.

The generation of the Hasse diagram of the restricted layout structure requires two functions. Initially, the structural objects have to be identified using the \texttt{itemlist} function. The function's arguments, \texttt{datadesign} and \texttt{randomfacsid}, as explained above, are given by:

\begin{verbatim}
BIBDWheat_objects <- itemlist(datadesign = BIBDWheat, 
                              randomfacsid = c(0, 0, 0))
\end{verbatim}

The call of the \texttt{itemlist} function creates a class \texttt{rls} object, and the structural objects can be viewed by printing this object. For illustrative purposes, the class \texttt{rls} object is printed below. However, to conserve space and avoid unnecessary repetition, these objects will not be printed in the subsequent examples. 

\begin{verbatim}
print(BIBDWheat_objects)
\end{verbatim}

\begin{verbatim}  
The names of all the structural objects to assist the input choices of the 
Hasse diagram of the Restricted Layout Structure (RLS) are: 
[1] "Mean"                                            
[2] "Blocks"                                          
[3] "Varieties"                                       
[4] "Blocks^Varieties"                               

and the matrix you need to fill in with the randomisation objects which 
are present in the RLS is: 
 
  All Structural Objects Randomisation Objects
1 "Mean"                 "Mean"               
2 "Blocks"               "NULL"               
3 "Varieties"            "NULL"               
4 "Blocks^Varieties"     "NULL"              
\end{verbatim}

The two-column matrix printed above is the matrix that needs to be amended to define the randomisation objects which will be present in the restricted layout structure. This matrix is also available as the \texttt{TransferObject} item in the list of items stored of the \texttt{BIBDWheat\_objects} class \texttt{rls} object. In the initial \texttt{TransferObject}, all entries in the second column, except for the mean, are set to "NULL." These entries require edits to specify the corresponding randomisation objects that are then passed to the \texttt{hasserls} function. Any randomisation objects left as "NULL" are excluded from the Hasse diagram of the restricted layout structure. In this particular example, all objects are retained and displayed in the final diagram.

\begin{verbatim}
IBDWheat_rls <- BIBDWheat_objects$TransferObject 
IBDWheat_rls[ , 2] <- IBDWheat_rls[ , 1]
IBDWheat_rls[4, 2] <- c("Plot[Block]")
\end{verbatim}

To illustrate the randomisation, a single randomisation arrow will be included in the Hasse diagram. This arrow represents the randomisation from the 'Varieties' randomisation object to the 'Plot within Block' randomisation object. To pass this information to \texttt{hasserls}, a two-column matrix is needed. In this case, the matrix will be of a single row, since there is only one randomisation arrow required. The first column entry represents the arrow's starting object index from \texttt{BIBDWheat\_objects}, and the second column entry represents the arrow's ending object index from \texttt{BIBDWheat\_objects}. From the printout of the \texttt{BIBDWheat\_objects} above, these are index numbers 3 and 4, respectively in this case.  

\begin{verbatim}
IBDWheat_rand_arrows <- matrix(c(3, 4), ncol = 2, byrow = TRUE) 
\end{verbatim}

Additionally, the \texttt{equation.out} option is set to "Y" (default is "N"), which generates a recommended mixed model equation based on the restricted layout structure. By including this feature, the \texttt{hasserls} function offers an algebraic expression of the model that can be used for statistical analysis. To get a shorter randomisation arrow than the default length, the argument \texttt{arrow.pos} is set to 10 (default is 7.5). For consistency, the font and font sizes are identical to the settings defined for the layout structure.

The Hasse diagram of the restricted layout structure based on the randomisation described above, is generated using the code below. The Hasse diagram is given in Figure~\ref{fig:bibdrls}.

\begin{verbatim}
hasserls(object = BIBDWheat_objects, randomisation.objects = IBDWheat_rls, 
         random.arrows = IBDWheat_rand_arrows, showRLS = "Y", 
         showpartialRLS = "Y", showdfRLS = "Y", showrandRLS = "Y",
         check.confound.df = "Y", maxlevels.df = "Y",  equation.out = "Y", 
         arrow.pos = 8.5, larger.fontlabelmultiplier = 1.8)
\end{verbatim}

The mixed model equation output is:
\begin{verbatim}
The suggested mixed model to be fitted is: 
 Response ~   as.factor(Blocks) +  as.factor(Varieties) +  
                as.factor(Blocks):as.factor(Varieties)
\end{verbatim}

\begin{figure}[H]
\centering
\includegraphics[width=\linewidth]{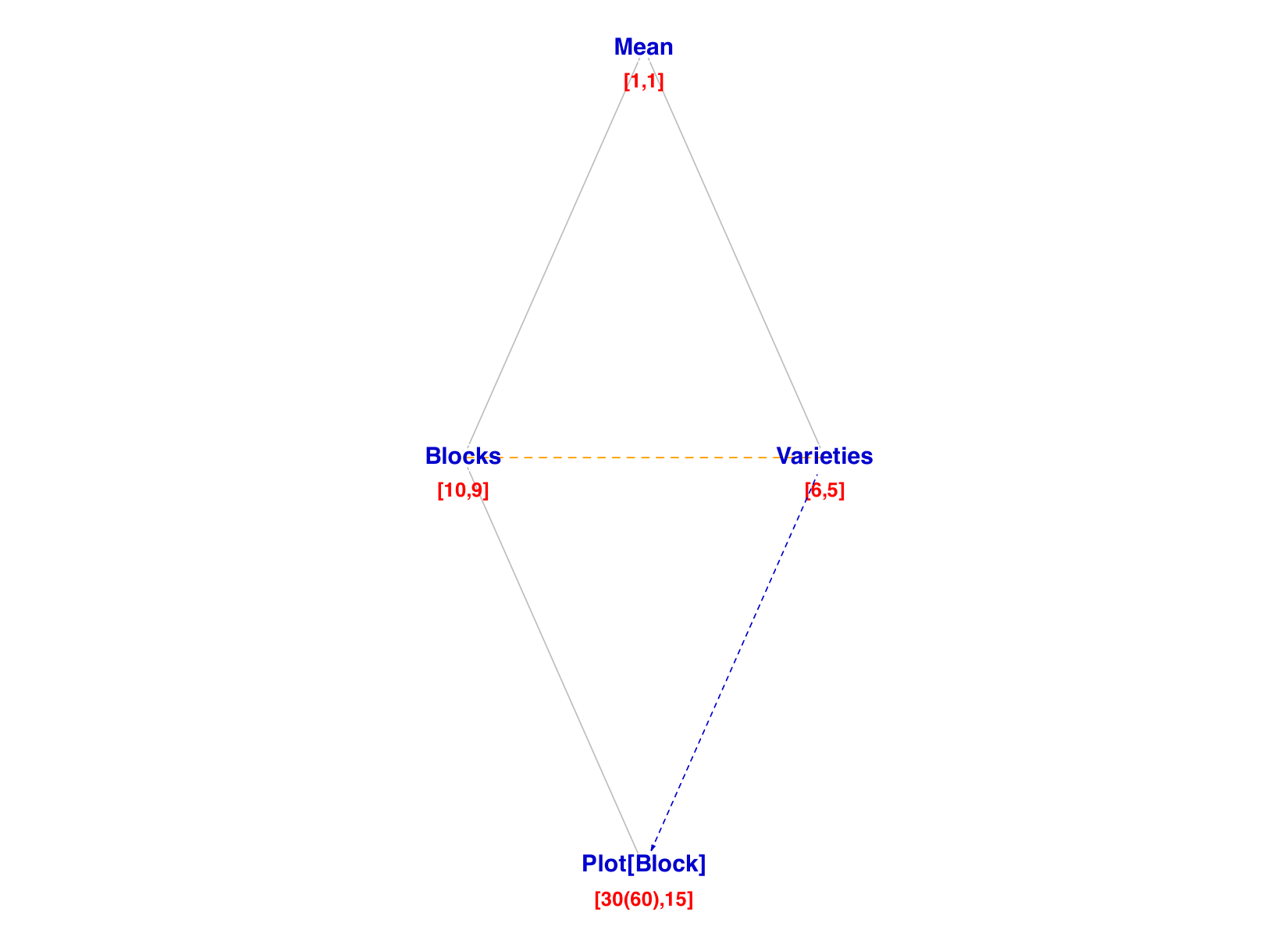}
\caption{Hasse diagram of the restricted layout structure for Example \ref{subsec:bibdexample}.}
\label{fig:bibdrls}
\end{figure}

By comparing the Hasse diagram of the layout structure in Figure~\ref{fig:bibdlayout} and the Hasse diagram of the restricted layout structure in Figure~\ref{fig:bibdrls}, the influence of the randomisation process on the experimental design can be visualised. This provides a valuable insight into the design's underlying structure.

\subsection{Example. Factorial design} \label{subsec:factorialexample}

\citet[p. 199]{box2005} describe a process development study, based on a factorial design, to investigate the effect of catalyst charge, temperature, pressure and reactant concentration on a chemical reaction. The four categorical factors included in the experimental design are Catalyst charge (levels: 10 and 15), Temperature (levels: 220 and 240), Pressure (levels: 50 and 80) and Concentration (levels: 10 and 12). 

The design used was a single replicate of a $2^{4}$ factorial design, and hence, involved 16 experimental runs. The design and response is available in the \texttt{Fac4Proc.dat} dataset of the \texttt{dae} package in \texttt{R}. The dataset is loaded via the following code:

\begin{verbatim}
data(Fac4Proc.dat)
\end{verbatim}

As in Ex. \ref{subsec:bibdexample}, it is essential to remove the response variable from the dataset before passing it to the functions in \texttt{hassediagrams}. The response variable is in the 6th column and is removed via:

\begin{verbatim}
Fact4Proc <- Fac4Proc.dat[ , -6]
\end{verbatim}

To generate the Hasse diagram of the layout structure, the \texttt{hasselayout} function is used where the single mandatory argument is the design studied, (stored in item \texttt{Fact4Proc}). 

For demonstration purposes and computational efficiency, the partial crossing between structural objects from the Hasse diagram of the layout structure is suppressed.

All factors of the design are fixed factors, so the \texttt{randomfacsid} is set to the NULL default option, which is equivalent to setting all factors as fixed. Other options, for example, \texttt{showdfLS} and \texttt{check.confound.df}, etc., remain at their default settings.

As there are more than five objects at the 3rd level in the Hasse diagram, the optional argument \texttt{smaller.fontlabelmultiplier} that controls the size of the font used in the label for these objects is set to 2.35 to increase the font from the default 1. The \texttt{larger.fontlabelmultiplier} is also set to 1.25.

\begin{figure}[H]
\centering
\includegraphics[width=\linewidth]{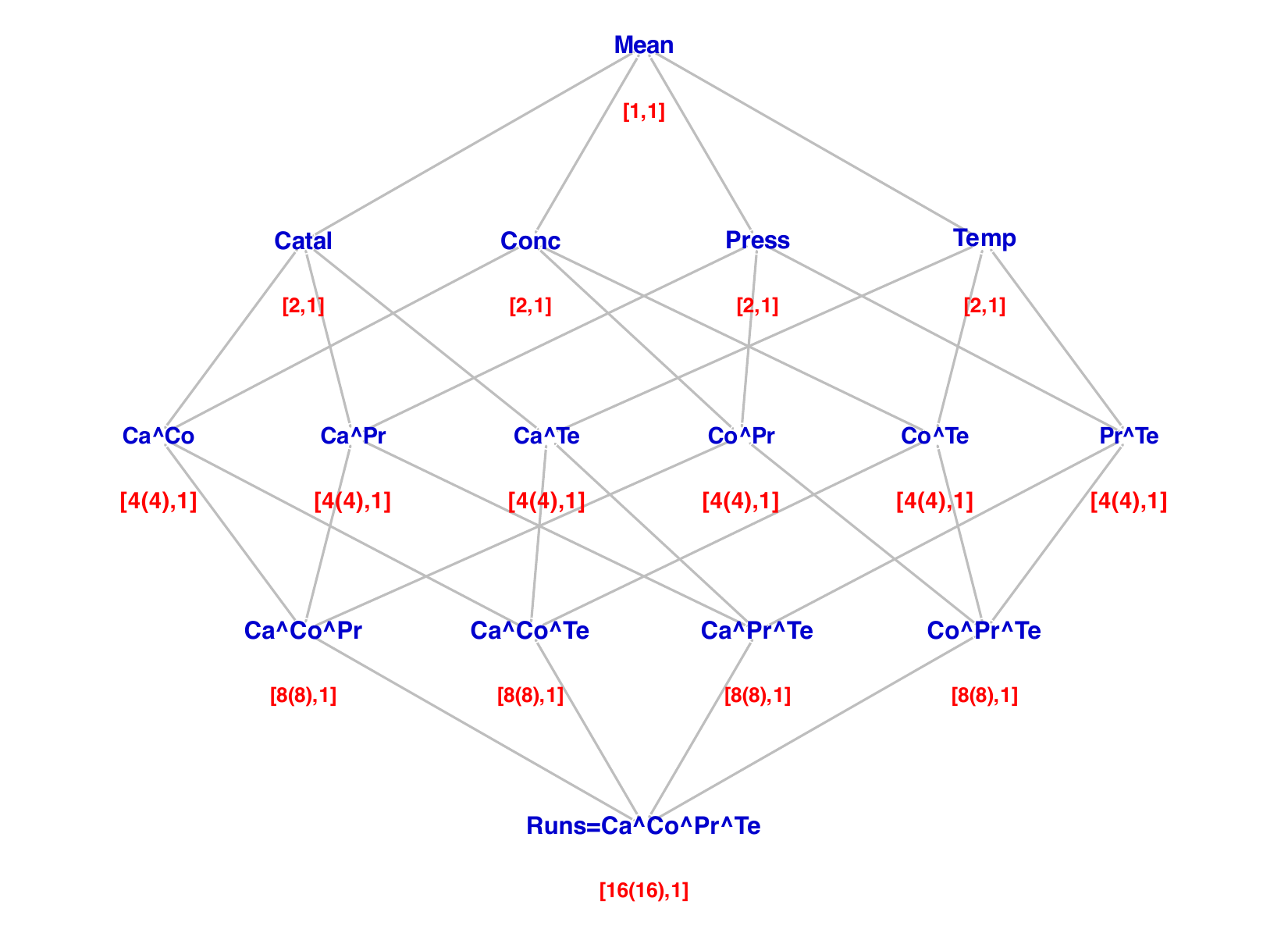}
\caption{Hasse diagram of the layout structure for Example \ref{subsec:factorialexample}.}
\label{fig:factorialdesign}
\end{figure}

The Hasse diagram of the layout structure of the factorial design is generated via the code below and the diagram is given in Figure~\ref{fig:factorialdesign}. Note that the generalised factor involving the four factors is equivalent to the Run factor (levels: 1 to 16). 

\begin{verbatim}
hasselayout(datadesign = Fact4Proc, showpartialLS = "N",
            larger.fontlabelmultiplier = 1.25,
            smaller.fontlabelmultiplier = 2.35)
\end{verbatim}


The randomisation performed was a conforming randomisation of a factorial design (see \citet{bate2016b}) where the run order was completely randomised. This is equivalent to independently randomising the levels of the four factors to the Run factor, i.e., 
\begin{equation}
\label{eq:runrandom}
Catalyst \land Temperature \land Pressure \land Concentration \to Run.
\end{equation}

This information can be used to generate the restricted layout structure. The first step is to create the class \texttt{rls} object using the \texttt{itemlist} function. 

The mandatory \texttt{datadesign} argument of the \texttt{itemlist} function was already set before to \texttt{Fact4Proc} in the layout structure and the \texttt{randomfacsid} argument can be its default setting. The \texttt{rls} object can be created via:

\begin{verbatim}
Fact4Proc_objects <- itemlist(datadesign = Fact4Proc)
\end{verbatim}

and the structural objects of the layout structure can be printed via:

\begin{verbatim}
summary(Fact4Proc_objects)
\end{verbatim}

The above print illustrates the two-column matrix that needs to be filled with the randomisation objects which will be present in the restricted layout structure (not shown here, but it is the equivalent print from Section \ref{subsec:bibdexample}). This matrix is the \texttt{TransferObject} item in the list of available items stored in the class \texttt{rls} object. 

All entries, except the mean, in the second column of the \texttt{TransferObject} are "NULL". The second step is to edit the second column of the \texttt{TransferObject} accordingly and pass it to the \texttt{hasserls} function. Any randomisation objects that remain as "NULL" will be removed. In this specific example, all objects are kept in the restricted layout structure. To reflect that the run order was completely randomised, as shown in \eqref{eq:runrandom}, the bottom randomisation object is renamed accordingly. 

\begin{verbatim}
Fact4Proc_rls <- Fact4Proc_objects$TransferObject 
Fact4Proc_rls[ , 2] <- Fact4Proc_rls[ , 1]
Fact4Proc_rls[16, 2] <- c("Catal^Conc^Press^Temp \u2192 Run")
\end{verbatim}

where \verb|\u2192| is the $\to$ symbol.

The final step to generate the Hasse diagram of the restricted layout structure is to build the \texttt{hasserls} function. The mandatory arguments \texttt{object} and \texttt{randomisation.objects} are the class \texttt{rls} object generated via \texttt{itemlist}, and the \texttt{Fact4Proc\_rls} two-column matrix edited using the \texttt{TransferObject} item. 

There are no randomisation arrows linking non-equivalent randomisation objects in this factorial design example, and therefore the \texttt{random.arrows} argument remains at its default setting (which is NULL). Similar to the layout structure, the partial crossing between randomisation objects is suppressed from the Hasse diagram by setting \texttt{showpartialRLS} to "N". Since one (or more) of the randomisation object labels contain Unicode characters, i.e., \verb|\u2192|, the font must be set to a Unicode-friendly font, for instance, "Arial Unicode MS", or write the plot as PDF. As in the layout structure, the \texttt{smaller.fontlabelmultiplier} is set to 2.35 and the \texttt{larger.fontlabelmultiplier} is set to 1.25.

\begin{figure}[H]
\centering
\includegraphics[width=\linewidth]{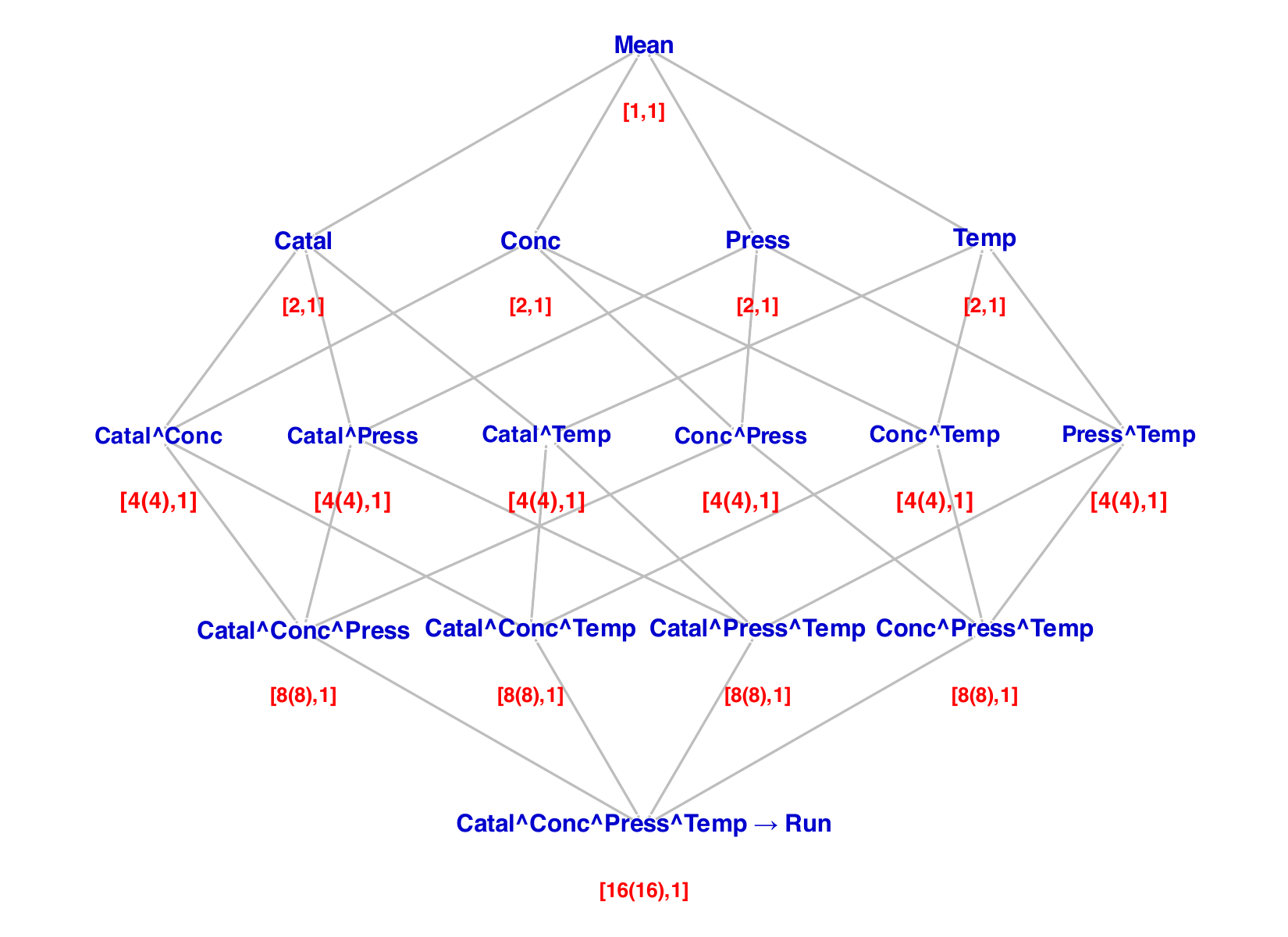}
\caption{Hasse diagram of the restricted layout structure for Example \ref{subsec:factorialexample}.}
\label{fig:factorialdesignrls}
\end{figure}

Finally, the Hasse diagram of the restricted layout structure of the factorial design based on randomisation discussed above can be generated via the following code:

\begin{verbatim}
hasserls(object = Fact4Proc_objects, randomisation.objects = Fact4Proc_rls,
         showpartialRLS = "N", smaller.fontlabelmultiplier = 2.35, 
         larger.fontlabelmultiplier = 1.25, hasse.font = "Arial Unicode MS")
\end{verbatim}

assuming the "Arial Unicode MS" font is available (other Unicode-friendly fonts can be used). The Hasse diagram of the restricted layout structure that is in Figure~\ref{fig:factorialdesignrls} is using the default "sans" font and generated via the \texttt{pdf} option such that:

\begin{verbatim}
hasserls(object = Fact4Proc_objects, randomisation.objects = Fact4Proc_rls,
         showpartialRLS = "N", smaller.fontlabelmultiplier = 2.35, 
         larger.fontlabelmultiplier = 1.25, pdf = "Y")
\end{verbatim}

If in addition to the Hasse diagram, it is of interest to print the table that represents the relationships between the randomisation objects in the restricted layout structure, then the \texttt{table.out} argument must be set to "Y". 

\subsection{Example. Crossover design} \label{subsec:crossoverexample}

In this section the dental crossover design using the dental dataset, which is built into the \texttt{hassediagrams} package, is assessed. A description of the crossover design and its factors is given in Section \ref{subsec:dental}. This built-in dataset is available as \texttt{dental}.

To begin with, the Hasse diagram of the layout structure is generated using the \texttt{hasselayout} function. The single mandatory argument is the design studied, which is built-in and stored under item \texttt{dental}. Out of the five factors, Subject is deemed to be a random factor and the remaining are fixed. This will be reflected in the code through the \texttt{randomfacsid} argument which will be a vector of length five, with all entries set at 0, i.e., fixed, except the second entry related to Subject, which will be set to 1, i.e., random. 

All remaining settings are kept to their default values, except for the argument \texttt{maxlevels.df}, which will be disabled. 

Disabling the \texttt{maxlevels.df} argument will suppress the inclusion of the potential maximum number of levels of each generalised factor in the structural object label, altering the appearance of the resulting Hasse diagram. This adjustment is made here to allow a comparison of the Hasse diagram's output against other examples where the \texttt{maxlevels.df} option is enabled, thereby highlighting the effect of this setting on the visualisation of the experimental structure. This setting can be particularly useful for users who wish to generate clearer and less cluttered Hasse diagrams, allowing for a sharper focus on the layout structure without the additional complexity of displaying the maximum potential number of levels of the generalised factors. 

\begin{figure}[H]
\centering
\includegraphics[width=\linewidth]{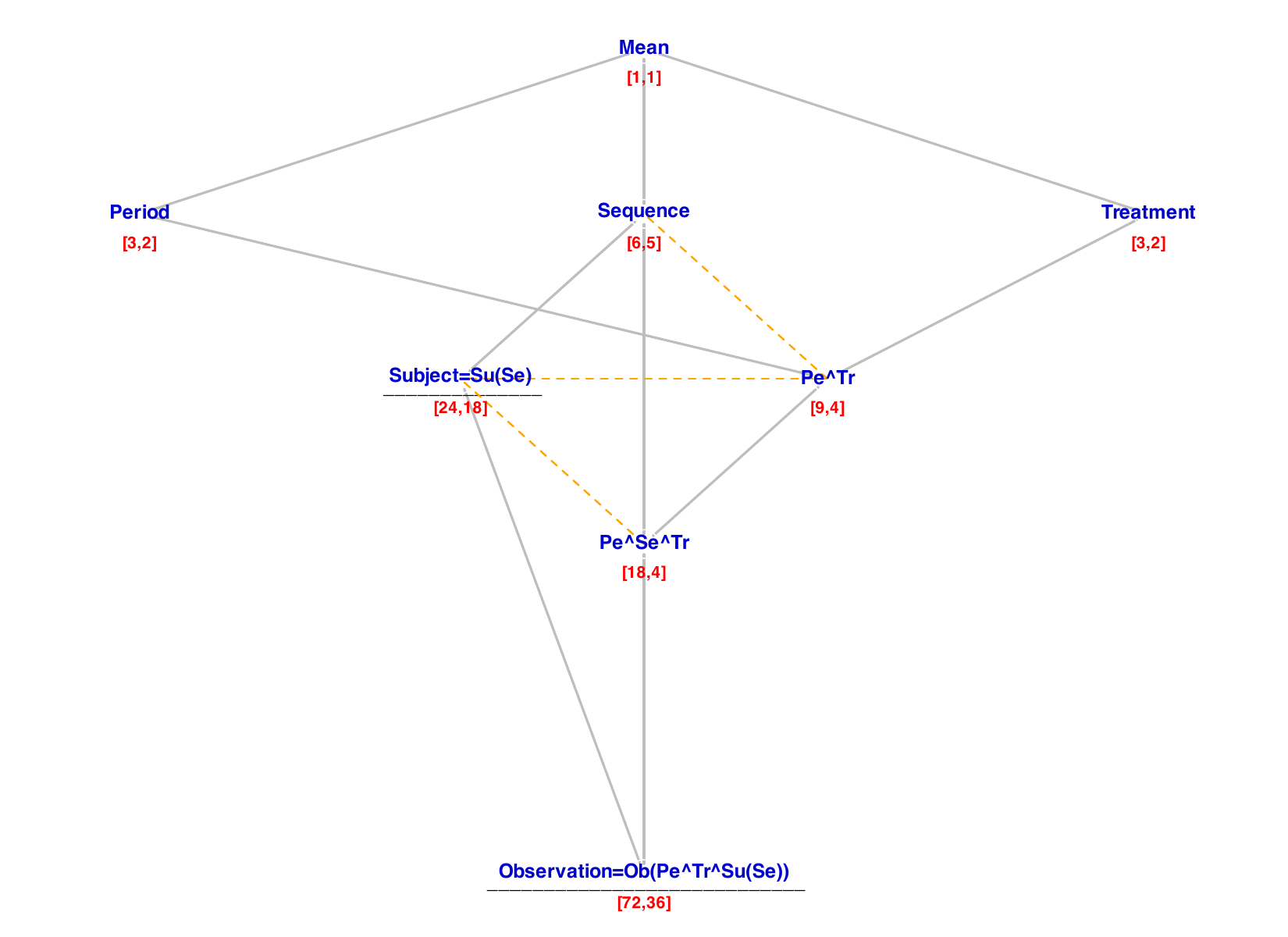}
\caption{Hasse diagram of the layout structure for Example \ref{subsec:crossoverexample}.}
\label{fig:crossoverlayout}
\end{figure}

The Hasse diagram of the layout structure of the \texttt{dental} crossover design is generated via the code below and is given in Figure~\ref{fig:crossoverlayout}.

\begin{verbatim}
hasselayout(datadesign = dental, randomfacsid = c(0, 1, 0, 0, 0), 
            maxlevels.df =  "N", larger.fontlabelmultiplier = 1.8)
\end{verbatim}

Note, 'Observation' was originally defined as a fixed factor, but is equivalent to the five-way generalised factor (that includes the random factor Subject) and therefore it will be declared random. A warning message is printed by the \texttt{hasselayout} call to inform the user of this automatic reclassification. The message advises the user to manually adjust the designation of the term if this reclassification is incorrect. 

The randomisation of the crossover design involves two stages and requires knowledge of the experimental scheme. The first stage involves randomly assigning the subjects to the treatment sequences. As a randomisation arrow never starts at the randomisation object that corresponds to a structural object in a lower level of the design, this randomisation is therefore defined as:

\begin{equation}
\label{eq:dentalrandom1}
Sequence \to Subject.
\end{equation}

For the second stage, the treatment labels are randomised by randomly assigning the treatments to the period by sequence combinations. In this case there is a structure to the treatment allocation, as documented in the experiment scheme. The treatment sequences are defined such that the design is balanced for first order carry-over effects \citep{jones2003} using Williams Double Latin squares \citep{bate2008}. For the randomisation to be defined as conforming \citep{bate2016b} the randomisation must respect these properties of the design, as given in the experiment scheme. Hence, it is not possible to permute the periods as this will potentially ruin the balance of the design. This is a constraint on the conforming randomisation. The conforming randomisation can still be defined using a randomisation arrow as treatments are randomised to combinations of period and sequence, hence
\begin{equation}
\label{eq:dentalrandom2}
Treatment \to Period \land Sequence.
\end{equation}

Using the rules described in Section \ref{sec:reslayout}, all structural objects in the layout structure will have equivalent randomisation objects in the restricted layout structure, with the exception of the $Period \land Sequence \land Treatment$ generalised factor. The randomisation objects in the restricted layout structure will be $Period$, $Sequence$, $Treatment$, $Subject$, $Period \land Sequence$ and $Observation$.

The first step to generate  the Hasse diagram of the restricted layout structure is to create the class \texttt{rls} object using the \texttt{itemlist} function. The \texttt{rls} object can be created via:

\begin{verbatim}
dental_objects <- itemlist(datadesign = dental, 
                           randomfacsid = c(0, 1, 0, 0, 0))
\end{verbatim}

and the structural objects of the layout structure can be printed via:

\begin{verbatim}
print(dental_objects)
\end{verbatim}

Once again, this illustrates the two-column matrix that needs to be filled with the randomisation objects which will be present in the restricted layout structure. This matrix is the \texttt{TransferObject} item in the list of available items stored in the class \texttt{rls} object. In this particular example, all objects, except the 6th object which corresponds to the Period by Treatment generalised factor, are retained and displayed in the final Hasse diagram of the restricted layout structure.

\begin{verbatim}
dental_rls <- dental_objects$TransferObject 
dental_rls[c(1:5, 7, 8), 2] <- c("Mean", "Period", "Sequence", 
                                 "Treatment", "Subject", 
                                 "Period^Sequence", 
                                 "Observation")
\end{verbatim}

To illustrate the randomisations, two randomisation arrows will be included in the Hasse diagram of the restricted layout structure. The first arrow represents the randomisation of the sequences to the subjects. The second arrow represents the randomisation of the treatments to the period by sequence combinations. To pass this information to \texttt{hasserls}, a two-column matrix is needed. In this case, the matrix will have two rows, one for each randomisation arrow. The first column entries represent the object indexes from \texttt{dental\_objects} at the start of the arrows; these are index numbers 3 and 5, respectively. The second column entries represent the object indexes from \texttt{dental\_objects} that the arrows point at; these are index numbers 4 and 7, respectively.

\begin{verbatim}
dental_rand_arrows <- matrix(c(3, 5, 4, 7), ncol = 2, byrow = TRUE) 
\end{verbatim}

\begin{figure}[H]
\centering
\includegraphics[width=\linewidth]{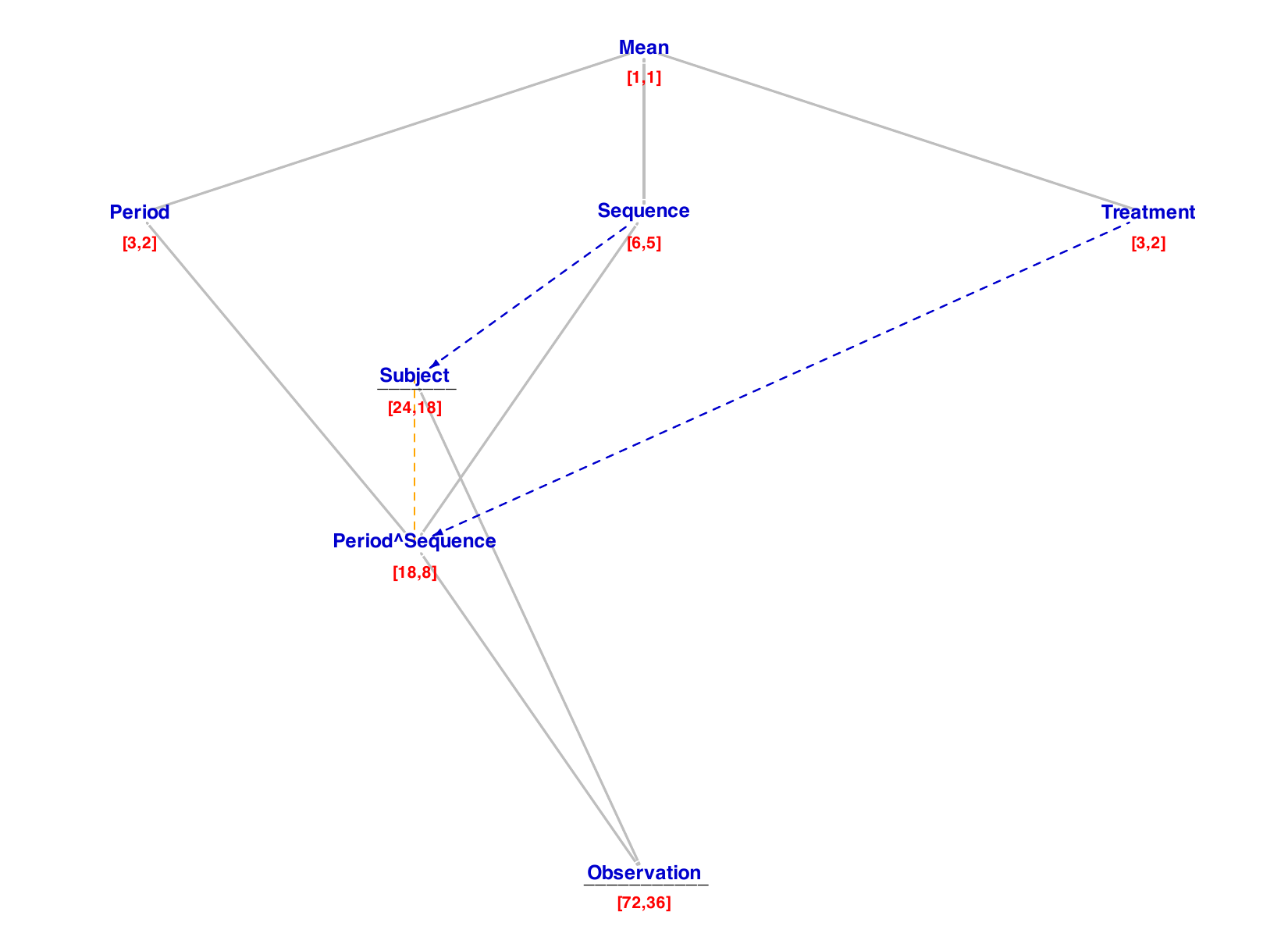}
\caption{Hasse diagram of the restricted layout structure for Example \ref{subsec:crossoverexample}.}
\label{fig:crossoverrls}
\end{figure}

Additionally, the \texttt{maxlevels.df} argument will remain disabled, consistent with the Hasse diagram of the layout structure, to suppress the display of potential maximum levels in the labels of the randomisation objects. Finally, the Hasse diagram of the restricted layout structure based on the randomisation described above is generated via the code below. The Hasse diagram is given in Figure~\ref{fig:crossoverrls}.

\begin{verbatim}
hasserls(object = dental_objects, randomisation.objects = dental_rls, 
         random.arrows = dental_rand_arrows, maxlevels.df = "N",
         larger.fontlabelmultiplier = 1.8)
\end{verbatim}

Through this example, the versatility of \texttt{hassediagrams} in handling complex experimental designs is demonstrated, particularly in crossover designs where the relationships between randomisation and structural factors play a crucial role in understanding the design's structure.

\subsection{Example. Split-plot in a row-column design} \label{subsec:splitplotexample}

\citet[Example 11]{brien2006} and \citet[p. 309–311]{cochran1957} describe an experiment involving a split-plot in a row-column design. The experiment aimed to investigate the effects of four soil treatments (factor: Soil), which were randomly assigned to four plants (factor: Plant) on each of three benches (factor: Bench) using a randomised complete block design. In addition to the soil treatments, three leaf treatments (factor: Treat) were applied to leaves (Leaf) located at the top, middle, or bottom layers of each plant (factor: Lyr).

The factors of interest in this experiment are Soil and Treat, while Bench, Plant, and Leaf are identified as 'inherent' factors \citep{bate2016a}. It was hypothesised that a systematic effect might occur down the plant, leading to the inclusion of the Lyr factor (levels: top, middle, and bottom) to account for the leaf position within the plant. Bench, Lyr, Soil, and Treat are fully crossed with one another, while Plant is inherently nested within Bench, and Leaf is nested within both Plant and Lyr. Hence Bench has three levels, Plant has twelve levels, and Leaf has thirty-six levels.

A simulated version of the design and treatment allocation, pre-randomisation, based on the description above is given in Table~\ref{table:splitplot}. The simulated design can be loaded as a data frame in \texttt{R} using the following code.

\begin{verbatim}
splitplot_doe <- data.frame(
    Bench = c(1, 1, 1, 1, 1, 1, 1, 1, 1, 1, 1, 1, 2, 2, 2, 2, 2, 2, 
              2, 2, 2, 2, 2, 2, 3, 3, 3, 3, 3, 3, 3, 3, 3, 3, 3, 3),
    Plant = c(1, 1, 1, 2, 2, 2, 3, 3, 3, 4, 4, 4, 5, 5, 5, 6, 6, 6, 7, 7,
              7, 8, 8, 8, 9, 9, 9, 10, 10, 10, 11, 11, 11, 12, 12, 12),
    Lyr = c("Top", "Middle", "Bottom", "Top", "Middle", "Bottom", 
            "Top", "Middle", "Bottom", "Top", "Middle", "Bottom",
            "Top", "Middle", "Bottom", "Top", "Middle", "Bottom", 
            "Top", "Middle", "Bottom", "Top", "Middle", "Bottom",
            "Top", "Middle", "Bottom", "Top", "Middle", "Bottom", 
            "Top", "Middle", "Bottom", "Top", "Middle", "Bottom"),
    Soil = c(3, 3, 3, 2, 2, 2, 1, 1, 1, 0, 0, 0, 0, 0, 0, 2, 2, 2, 
             1, 1, 1, 3, 3, 3, 3, 3, 3, 0, 0, 0, 2, 2, 2, 1, 1, 1),
    Treat = c(2, 0, 1, 1, 0, 2, 0, 1, 2, 1, 0, 2, 0, 2, 1, 0, 2, 1, 
              1, 2, 0, 1, 2, 0, 0, 1, 2, 2, 1, 0, 2, 1, 0, 2, 0, 1),
    Leaf = 1:36
)
\end{verbatim}

\begin{table}[H]
\centering
\begin{tabular}{llllll}
\hline
Bench & Plant & Lyr    & Soil & Treat & Leaf \\
\hline
1     & 1     & Top    & 3    & 2     & 1    \\
1     & 1     & Middle & 3    & 0     & 2    \\
1     & 1     & Bottom & 3    & 1     & 3    \\
1     & 2     & Top    & 2    & 1     & 4    \\
1     & 2     & Middle & 2    & 0     & 5    \\
1     & 2     & Bottom & 2    & 2     & 6    \\
1     & 3     & Top    & 1    & 0     & 7    \\
1     & 3     & Middle & 1    & 1     & 8    \\
1     & 3     & Bottom & 1    & 2     & 9    \\
1     & 4     & Top    & 0    & 1     & 10   \\
1     & 4     & Middle & 0    & 0     & 11   \\
1     & 4     & Bottom & 0    & 2     & 12   \\
2     & 5     & Top    & 0    & 0     & 13   \\
2     & 5     & Middle & 0    & 2     & 14   \\
2     & 5     & Bottom & 0    & 1     & 15   \\
2     & 6     & Top    & 2    & 0     & 16   \\
2     & 6     & Middle & 2    & 2     & 17   \\
2     & 6     & Bottom & 2    & 1     & 18   \\
2     & 7     & Top    & 1    & 1     & 19   \\
2     & 7     & Middle & 1    & 2     & 20   \\
2     & 7     & Bottom & 1    & 0     & 21   \\
2     & 8     & Top    & 3    & 1     & 22   \\
2     & 8     & Middle & 3    & 2     & 23   \\
2     & 8     & Bottom & 3    & 0     & 24   \\
3     & 9     & Top    & 3    & 0     & 25   \\
3     & 9     & Middle & 3    & 1     & 26   \\
3     & 9     & Bottom & 3    & 2     & 27   \\
3     & 10    & Top    & 0    & 2     & 28   \\
3     & 10    & Middle & 0    & 1     & 29   \\
3     & 10    & Bottom & 0    & 0     & 30   \\
3     & 11    & Top    & 2    & 2     & 31   \\
3     & 11    & Middle & 2    & 1     & 32   \\
3     & 11    & Bottom & 2    & 0     & 33   \\
3     & 12    & Top    & 1    & 2     & 34   \\
3     & 12    & Middle & 1    & 0     & 35   \\
3     & 12    & Bottom & 1    & 1     & 36   \\
\hline
\end{tabular}
\caption{The full split-plot in a row-column design of Section \ref{subsec:splitplotexample}.}
\label{table:splitplot}
\end{table}

The factors Bench, Plant, Lyr, and Leaf are random, and the factors Soil and Treat are fixed. These will be reflected in the \texttt{randomfacsid} argument in \texttt{hasselayout} and \texttt{itemlist}.

For the layout structure, the dotted lines indicating partial crossing of objects are excluded for clarity by setting the \texttt{showpartialLS} argument to "N". For demonstration purpose of the option, the black and white option Hasse diagram is enabled by setting the \texttt{produceBWplot} argument to "Y". On top of the generation of the Hasse diagram of the layout structure, an interim layout structure table that shows the relationships between the structural objects in the layout structure \citep{bate2016a} is generated using the \texttt{table.out} argument is set to "Y". The small font multiplier is set to 3.3, the large font multiplier is set to 1.8, and the middle font multiplier for the generalised factor is set to 2.3.

\begin{figure}[H]
\centering
\includegraphics[width=\linewidth]{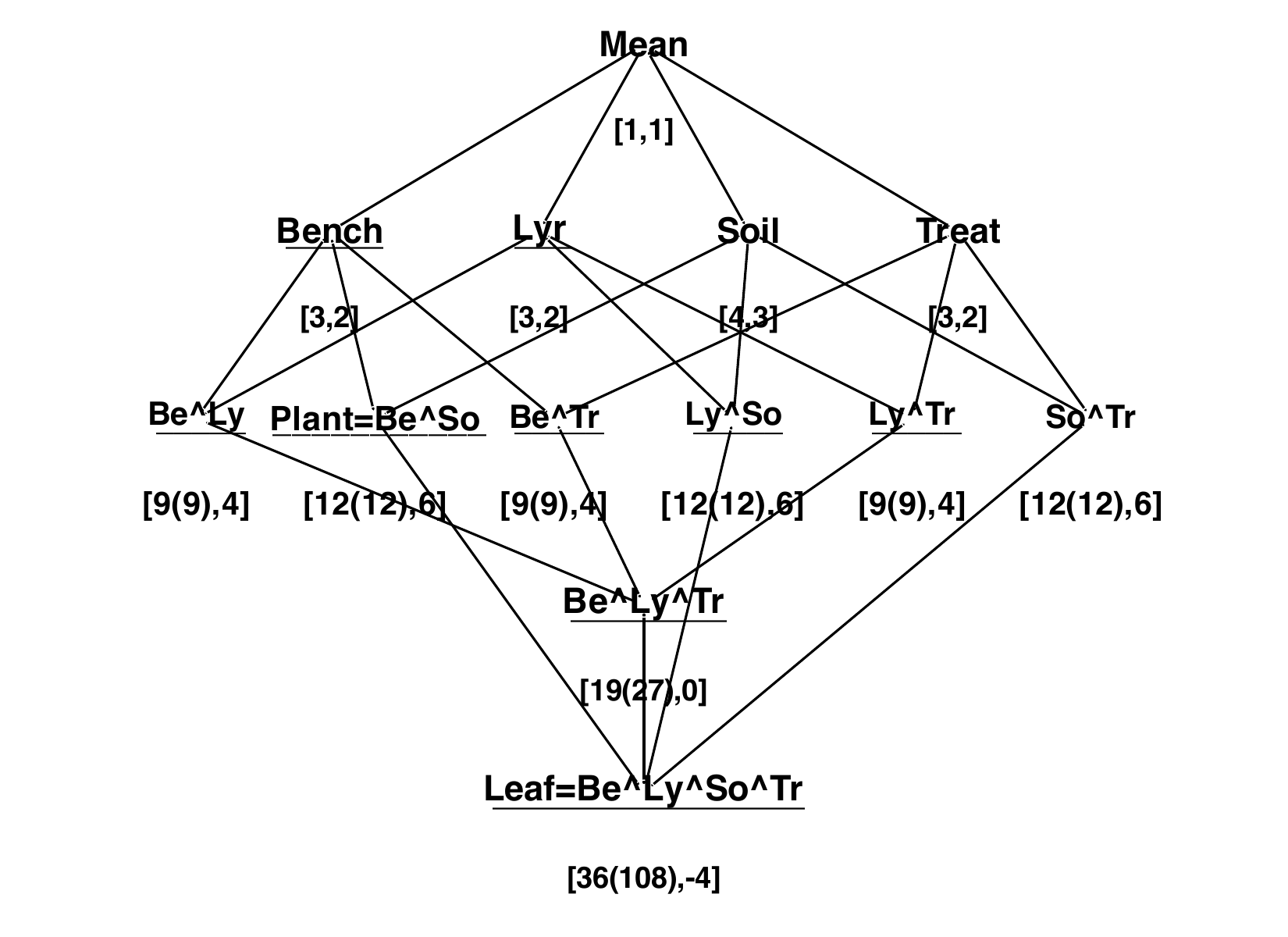}
\caption{Hasse diagram of the layout structure for Example \ref{subsec:splitplotexample}.}
\label{fig:spls}
\end{figure}

The Hasse diagram of the layout structure of the split-plot design is generated via the code below and the diagram is given in Figure~\ref{fig:spls}. 

\begin{verbatim}
hasselayout(data=splitplot_doe, randomfacsid=c(1, 1, 1, 0, 0, 1),
            showpartialLS = "N", table.out = "Y", produceBWPlot = "Y", 
            smaller.fontlabelmultiplier = 3.3, 
            larger.fontlabelmultiplier = 1.8,
            middle.fontlabelmultiplier = 2.3)
\end{verbatim}

\vspace{2cm}

{\notsotiny
\begin{verbatim}
The following table shows the relationships between the factors and generalised factors in the Layout Structure
                     Mean Bench Lyr   Soil  Treat Be^Ly Plant=Be^So Be^Tr Ly^So Ly^Tr So^Tr Be^Ly^Tr Leaf=Be^Ly^So^Tr
Mean                 " "  "(0)" "(0)" "(0)" "(0)" "(0)" "(0)"       "(0)" "(0)" "(0)" "(0)" "(0)"    "(0)"           
Bench                "1"  " "   "0"   "0"   "0"   "(0)" "(0)"       "(0)" "0"   "(0)" "0"   "(0)"    "(0)"           
Lyr                  "1"  "0"   " "   "0"   "0"   "(0)" "0"         "(0)" "(0)" "(0)" "0"   "(0)"    "(0)"           
Soil                 "1"  "0"   "0"   " "   "0"   "0"   "(0)"       "0"   "(0)" "0"   "(0)" "(0)"    "(0)"           
Treat                "1"  "0"   "0"   "0"   " "   "(0)" "0"         "(0)" "0"   "(0)" "(0)" "(0)"    "(0)"           
Bench^Lyr            "1"  "1"   "1"   "0"   "(0)" " "   "(0)"       "(0)" "(0)" "(0)" "(0)" "(0)"    "(0)"           
Bench^Soil           "1"  "1"   "0"   "1"   "0"   "(0)" " "         "(0)" "(0)" "(0)" "(0)" "(0)"    "(0)"           
Bench^Treat          "1"  "1"   "(0)" "0"   "1"   "(0)" "(0)"       " "   "(0)" "(0)" "(0)" "(0)"    "(0)"           
Lyr^Soil             "1"  "0"   "1"   "1"   "0"   "(0)" "(0)"       "(0)" " "   "(0)" "(0)" "(0)"    "(0)"           
Lyr^Treat            "1"  "(0)" "1"   "0"   "1"   "(0)" "(0)"       "(0)" "(0)" " "   "(0)" "(0)"    "(0)"           
Soil^Treat           "1"  "0"   "0"   "1"   "1"   "(0)" "(0)"       "(0)" "(0)" "(0)" " "   "(0)"    "(0)"           
Bench^Lyr^Treat      "1"  "1"   "1"   "(0)" "1"   "1"   "(0)"       "1"   "(0)" "1"   "(0)" " "      "(0)"           
Bench^Lyr^Soil^Treat "1"  "1"   "1"   "1"   "1"   "1"   "1"         "1"   "1"   "1"   "1"   "1"      " "             
\end{verbatim}
}

The interim layout structure table suggests that Bench$\land $Soil is equivalent to Plant and only one of the four three-way generalised factors (Bench$\land$Lyr$\land$Treat) is not nested within the remaining order 1 term (the other three are, and thus are equivalent to the four-way generalised factor).

Furthermore, the user is informed by the program that there are six confounded degrees of freedom, and hence, some of the values presented in the diagram are unreliable. The informative message and a table of the structural objects with a description of the associated degrees of freedom generated by the \texttt{hasselayout} function is:

\begin{Verbatim}
There are 6 confounded degrees of freedom
                     Actual levels DF by Subtraction Potential Confounded DF
Bench                            3                 2                      No
Lyr                              3                 2                      No
Soil                             4                 3                      No
Treat                            3                 2                      No
Bench^Lyr                        9                 4                     Yes
Bench^Soil                      12                 6                     Yes
Bench^Treat                      9                 4                     Yes
Lyr^Soil                        12                 6                     Yes
Lyr^Treat                        9                 4                     Yes
Soil^Treat                      12                 6                     Yes
Bench^Lyr^Treat                 19                 0                     Yes
Bench^Lyr^Soil^Treat            36                -4                      No       
\end{Verbatim}

In this output each of the structural objects in the layout structure are listed. The columns correspond to the (i) object name, (ii) number of levels of the object present in the design, (iii) the number of degrees of freedom (calculated using the subtraction technique) and (iv) whether the objects share one or more degrees of freedom with another object. It can be that multiple objects share a single degree of freedom. The user should investigate where there are confounded degrees of freedom as the numbers calculated by the subtraction method will not be accurate.

Confounded degrees of freedom can indicate that the levels of the factors are not specified appropriately or that the design needs to be changed. However, sometimes once appropriate randomisation is performed, the layout structure modified to account for the post-randomised allocation of the factor levels, will no longer contain confounded degrees of freedom. 

After investigating where the confounded degrees of freedom occur the next step is to proceed to randomisation and the restricted layout structure. Similar to previous examples, the class \texttt{rls} object is generated via:

\begin{verbatim}
sp_objects <- itemlist(datadesign = splitplot_doe, 
                       randomfacsid=c(1, 1, 1, 0, 0, 1))
summary(sp_objects)
\end{verbatim}

The class \texttt{rls} object above is used to choose the randomisation objects for the restricted layout structure. 

The design was randomised using a conforming randomisation \citep{bate2016a}. To begin with, the soils were randomly assigned to the plants. This was performed separately for each bench, and hence, the randomisation can be defined as:

\begin{equation}
Soil \to Plant[Bench],
\end{equation}

and therefore, the first randomisation arrow goes from Soil to Plant[Bench] which are index numbers 4 and 7, for arrow's starting and pointing object indexes, respectively.

The leaf treatments were then randomly assigned to the layers on the plant by randomising the treatments to the plant layers across benches, by independently randomising the layers and benches, separately for those plants receiving each soil treatment. Hence, the second randomisation is dependent on the outcome of the first randomisation and is defined as incoherent by \citet{brien2006}. The randomisation can therefore be defined as:

\begin{equation}
Treat \to \{Bench \otimes Lyr\}[Soil],
\end{equation}

and therefore, the second randomisation arrow goes from Treat to \{Bench $\otimes$ Lyr\}[Soil] which are index numbers 5 and 13, for arrow's starting and pointing object indexes, respectively. Both randomisation arrows are specified to be passed to \texttt{hasserls} with the code below.

\begin{verbatim}
sp_rand_arrows <- matrix(c(4, 7, 5, 13), ncol = 2, byrow = TRUE) 
\end{verbatim}

The randomisation structure implies that the randomisation objects Bench, Soil Treat, Plant[Bench] Lyr[Soil], Soil$\land$Treat and \{Bench $\otimes$ Lyr\}[Soil] define the restricted layout structure.

\begin{verbatim}
sp_rls <- sp_objects$TransferObject
sp_rls[, 2] <- c("Mean", "Bench", "NULL", "Soil", "Treat", 
                 "NULL", "Plant[Bench]", "NULL", 
                 "Lyr[Soil]", "NULL", "Soil^Treat", 
                 "NULL", "Leaf={Bench \u2297 Lyr}[Soil]") 
\end{verbatim}

\begin{figure}[H]
\centering
\includegraphics[width=\linewidth]{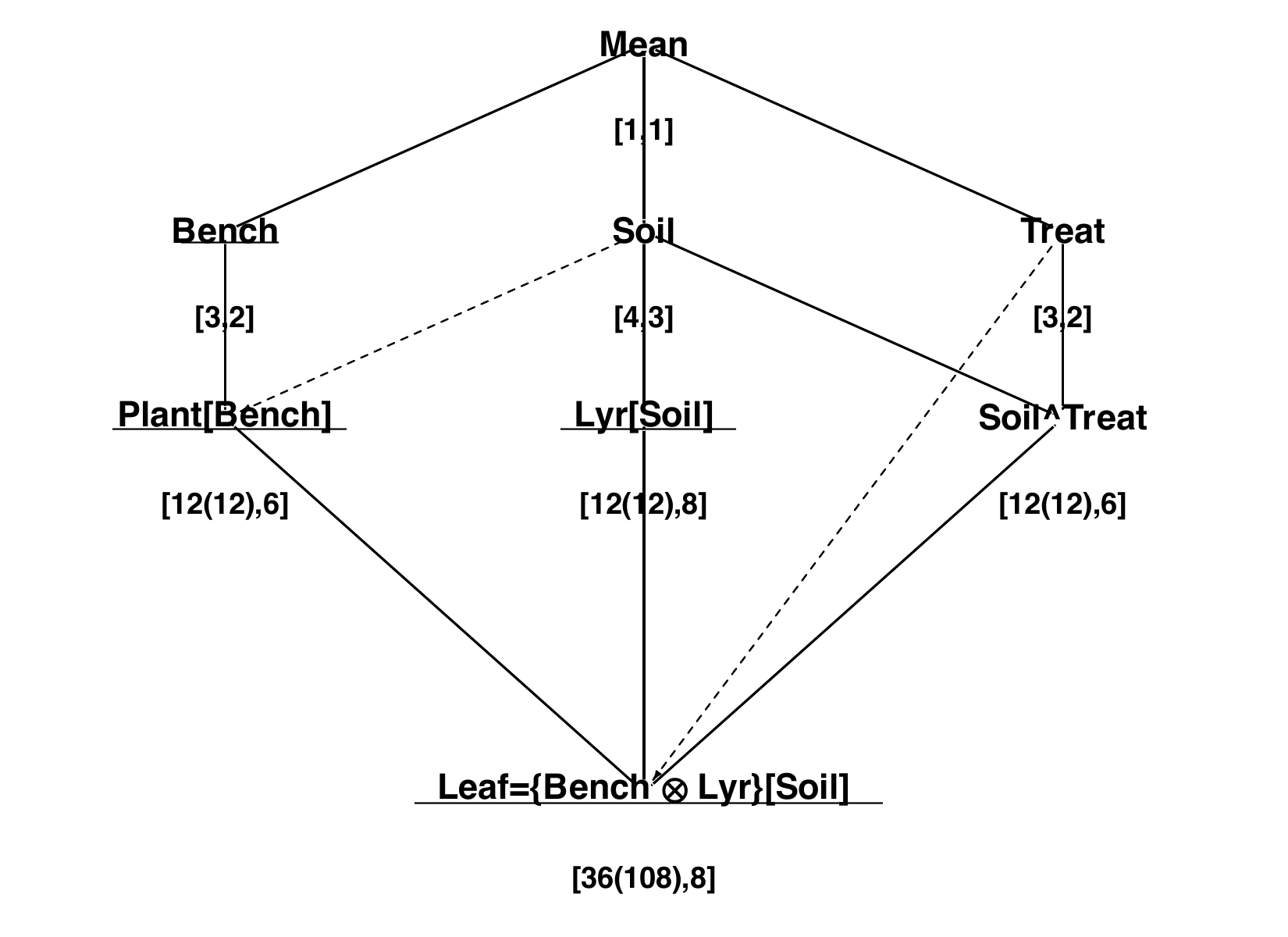}
\caption{Hasse diagram of the restricted layout structure for Example \ref{subsec:splitplotexample}.}
\label{fig:sprls}
\end{figure}

Finally, the Hasse diagram of the restricted layout structure, based on the randomisation described above, is generated via the code:

\begin{verbatim}
hasserls(object = sp_objects, randomisation.objects = sp_rls, 
         random.arrows = sp_rand_arrows, showpartialRLS = "N",
         produceBWPlot = "Y", hasse.font = "Arial Unicode MS",
         smaller.fontlabelmultiplier = 3.3, 
         larger.fontlabelmultiplier = 1.8,
         middle.fontlabelmultiplier = 2.3)
\end{verbatim}

assuming the "Arial Unicode MS" font is available (other Unicode-friendly fonts can be used). The Hasse diagram of the restricted layout structure that is in Figure~\ref{fig:sprls} is using the default "sans" font and generated via the \texttt{pdf} option such that:

\begin{verbatim}
hasserls(object = sp_objects, randomisation.objects = sp_rls, 
         random.arrows = sp_rand_arrows, showpartialRLS = "N",
         produceBWPlot = "Y", pdf = "Y",
         smaller.fontlabelmultiplier = 3.3, 
         larger.fontlabelmultiplier = 1.8,
         middle.fontlabelmultiplier = 2.3)
\end{verbatim}

\section{Conclusion} \label{sec:conclusion}

The development of the \texttt{R} package \texttt{hassediagrams} contributes to the statistical software literature by providing a powerful tool for visualising the structural hierarchy in experimental designs, specifically through Hasse diagrams of layout and restricted layout structures. The methodology proposed by \citet{bate2016a} and \citet{bate2016b} is implemented in \texttt{hassediagrams}, enabling users to systematically identify structural relationships and incorporate randomisation effects within these designs, enhancing clarity in model selection and statistical inference.

The core functionality for generating layout and restricted layout Hasse diagrams are implemented by the \texttt{hasselayout}, \texttt{itemlist}, and \texttt{hasserls} functions within the package. The Hasse diagram of the restricted layout structure includes visualisation features to highlight the influence of the randomisation structure. In summary, the package leverages \texttt{R}’s graphical capabilities to deliver diagrams that effectively reveal potential weaknesses and dependencies among experimental factors pre- and post-randomisation.

Practical applications of \texttt{hassediagrams} are demonstrated through four examples, covering several types of experimental designs (factorial, BIBD, crossover, and split-plot), showcasing its capacity for revealing hierarchical structures and randomisation dependencies. Although the examples in this paper focus on basic experimental layouts for clear illustration, \texttt{hassediagrams} is well-suited for more complicated designs and can offer significant benefits for researchers managing complex structural relationships.

\section*{Disclaimer} \label{sec:disclaimer}

The content of this article are reflective of the authors own personal opinions and not necessarily those of GSK or other affiliated institutions.

\bibliographystyle{apalike} 
\bibliography{refs}

\end{document}